\documentclass[
reprint,
amsmath,
amssymb,
prb,
superscriptaddress,
nofootinbib,
notitlepage
]{revtex4-2}

\usepackage[colorlinks, breaklinks=true, linkcolor=blue, citecolor=blue, linktocpage=true]{hyperref}
\usepackage{color}
\usepackage{graphicx}
\usepackage{dcolumn}
\usepackage{bm}

\usepackage{txfonts}

\begin{document}

\title{Microwave-to-Optical Quantum Transduction with Antiferromagnets}

\author{Akihiko Sekine}
\email{akihiko.sekine@fujitsu.com}
\affiliation{Fujitsu Research, Fujitsu Limited, Kawasaki 211-8588, Japan}
\author{Ryo Murakami}
\affiliation{Fujitsu Research, Fujitsu Limited, Kawasaki 211-8588, Japan}
\author{Yoshiyasu Doi}
\affiliation{Fujitsu Research, Fujitsu Limited, Kawasaki 211-8588, Japan}

\date{\today}

\begin{abstract}
The quantum transduction, or equivalently quantum frequency conversion, between microwave and optical photons is essential for realizing scalable quantum computers with superconducting qubits.
Due to the large frequency difference between microwave and optical ranges, the transduction needs to be done via intermediate bosonic modes or nonlinear processes.
Regarding the transduction mediated by magnons, previous studies have so far utilized ferromagnetic magnons in ferromagnets.
Here, we formulate a theory for the microwave-to-optical quantum transduction mediated by antiferromagnetic magnons in antiferromagnets.
We derive analytical expressions for the transduction efficiency in the cases with and without an optical cavity, where a microwave cavity is used in both cases.
In contrast to the case of the quantum transduction using ferromagnets, we find that the quantum transduction can occur even in the absence of an external static magnetic field.
We also find that, in the case with an optical cavity the transduction efficiency takes a peak structure with respect to the sample thickness, indicating that there exists an optimal thickness, whereas in the case without an optical cavity the transduction efficiency is a monotonically increasing function of the sample thickness.
Our study opens up a way for possible applications of antiferromagnetic materials in future quantum interconnects.
\\
\\
\end{abstract}

\maketitle

\section{Introduction}
Antiferromagnets have attracted attention in the field of spintronics \cite{Baltz2018}.
Unlike ferromagnets, antiferromagnets do not generate unwanted stray fields because of the zero net magnetization, which can be a merit of utilizing antiferromagnets near the magnetic-field sensitive devices such as superconducting qubits.
Moreover, since the dynamics of antiferromagnets such as the antiferromagnetic resonance is typically in the terahertz regime, fast manipulation and utilization of antiferromagnet-based devices are expected.
For example, it has been suggested that antiferromagnets can be active elements of a memory \cite{Marti2014} and logic device \cite{Wadley2016}, complementing or replacing ferromagnets.

The quantum transduction, or quantum frequency conversion, between photons at different frequencies is an important quantum technology that enables the interconnects between distant quantum devices.
In particular, the quantum transduction between microwave and optical photons is vital for the realization of large-scale quantum computers with superconducting qubits \cite{Lauk2020,Lambert2020,Han2021}.
This is because, optical fibers at the telecom frequency $\approx 200\, \mathrm{THz}$ are suitable for long-range quantum state transfer with low loss even at room temperature, while microwaves, by which the superconducting qubits are manipulated, are not suitable for such a purpose.
Due to the large frequency difference between the microwave and optical ranges, the quantum transduction needs to be done via intermediate interaction processes between photons and a bosonic mode or via nonlinear interaction processes between photons, such as the optomechanical effect \cite{Andrews2014,Vainsencher2016,Higginbotham2018,Forsch2020,Mirhosseini2020,Jiang2020,Han2020,Arnold2020,Hoenl2022,Barzanjeh2022} electro-optic effect \cite{Rueda2016,Fan2018,Hease2020,Holzgrafe2020,McKenna2020,Xu2021,Youssefi2021,Sahu2022}, and magneto-optic effect \cite{Hisatomi2016,Osada2016,Zhang2016,Haigh2016,Zhu2020,Sekine2024}.

The microwave-to-optical quantum transduction via the magneto-optic Faraday effect, i.e., the light-magnon interaction have so far been studied in ferromagnets such as YIG \cite{Hisatomi2016,Osada2016,Zhang2016,Haigh2016,Zhu2020,Sekine2024}.
Such a quantum transduction mediated by ferromagnetic magnons can have a wide bandwidth and can be operated even at room temperature, although there is a room for improvement of the transduction efficiency that is low compared to other transduction methods \cite{Lauk2020,Lambert2020,Han2021}.
Also, the coherent coupling between a ferromagnetic magnon and a superconducting qubit, which is an important step toward the realization of the quantum state transfer between superconducting qubits via optical photon, has been realized \cite{Tabuchi2015,Viennot2015,Rameshti2022}.

To the best of our knowledge, the microwave-to-optical quantum transduction utilizing antiferromagnets has not yet been considered.
However, a recent experimental study has demonstrated a coherent coupling between microwave cavity photons and antiferromagnetic magnons \cite{Johansen2018,Boventer2023}.
Also, a recent theoretical work has studied the coupling between optical cavity photons and antiferromagnetic magnons \cite{Parvini2020}.
Therefore, we believe that the necessary ingredients for the microwave-to-optical quantum transduction utilizing antiferromagnets are ready to be put together.

In this paper, we study theoretically the quantum transduction between the microwave and optical frequency ranges mediated by antiferromagnetic magnons in antiferromagnets.
We derive analytical expressions for the transduction efficiency in the cases with and without an optical cavity, where a microwave cavity is used in both cases.
We estimate the magnitude of the transduction efficiency by substituting possible values of the material parameters and the cavity parameters.
We also investigate the thickness dependence of the transduction efficiency and find a distinct behavior of the transduction efficiency as a function of the thickness depending on the presence or absence of an optical cavity.

This paper is organized as follows.
In Sec.~\ref{Sec-Model} we introduce a generic lattice model for antiferromagnets with two sublattices and express it in terms of the antiferromagnetic magnon operators.
In Sec.~\ref{Sec-Microwave-Magnon} we derive the interaction between microwave cavity photons and antiferromagnetic magnons from the Zeeman interaction in antiferromagnets.
In Sec.~\ref{Sec-Light-Magnon} we derive the interaction between optical cavity photons and antiferromagnetic magnons in the case with an optical cavity, and the interaction between optical itinerant photons and antiferromagnetic magnons in the case without an optical cavity, both from a linear magneto-optic (Faraday) effect in antiferromagnets.
In Sec.~\ref{Sec-Quantum-Transduction}, starting from a generic description for calculating the transduction efficiency with the input-output formalism, we derive analytical expressions for the transduction efficiency of the microwave-to-optical quantum transduction in the cases with and without an optical cavity.
In Sec.~\ref{Sec-Numerical-Results} we present the numerical results for the transduction efficiency.
In Sec.~\ref{Sec-Discussion} we briefly discuss a possible experimental realization and the merits of our proposal utilizing antiferromagnets.
We also discuss possible ways to improve the transduction efficiency.
In Sec.~\ref{Sec-Summary} we summarize this study.

\section{Theoretical Model \label{Sec-Model}}
\begin{figure}[!b]
\centering
\includegraphics[width=\columnwidth]{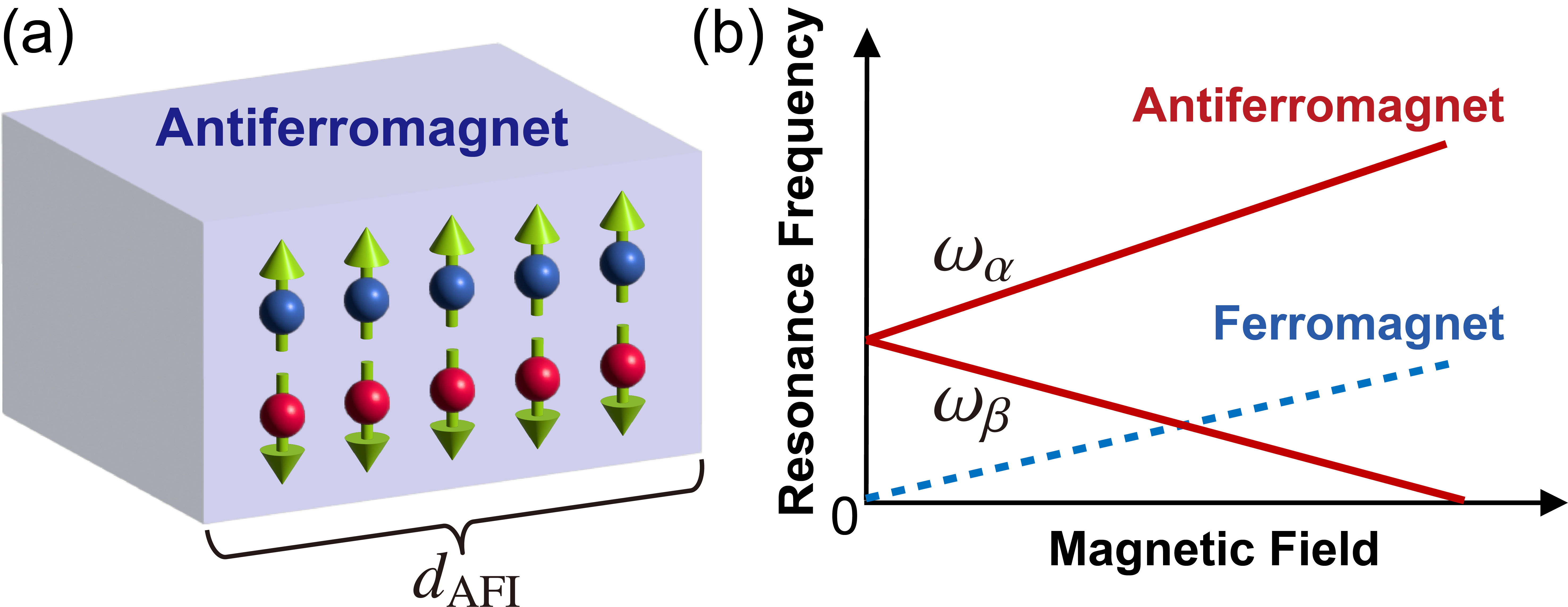}
\caption{Schematic illustration of (a) an antiferromagnet with two sublattices of thickness $d_{\mathrm{AFI}}$ and (b) the resonance frequencies of an antiferromagnet ($\omega_\alpha\ge \omega_\beta$) and that of a ferromagnet.
}\label{Fig1}
\end{figure}
We study the three-dimensional antiferromagnetic insulators with two sublattices $A$ and $B$ [see Fig.~\ref{Fig1}(a)], whose Hamiltonian is described by
\begin{align}
H_{\mathrm{AFM}}=&\ J\sum_{\langle i,j\rangle}\bm{S}_i\cdot\bm{S}_j+|\gamma|\sum_i\bm{B}_0\cdot\bm{S}_i-\frac{K_\parallel}{2}\sum_i\left(S_i^z\right)^2\nonumber\\
&+\frac{K_\perp}{2}\sum_i\left(S_i^x\right)^2,
\label{Original-Hamiltonian-AFM}
\end{align}
where the first through fourth terms are the exchange interaction with $J>0$ between nearest-neighbor spins, the Zeeman interaction with $\bm{B}_0=B_0\bm{e}_z$ being an external magnetic field and $|\gamma|$ being is the gyromagnetic ratio, the easy-axis anisotropy with $K_\parallel>0$, the hard-axis anisotropy with $K_\perp\ge 0$, respectively.

To quadratic order in the sublattice magnon operators $\hat{a}_{\bm{k}}$ and $\hat{b}_{\bm{k}}$ in the Holstein-Primakoff transformation \cite{Holstein1940} which satisfy the commutation relations $[\hat{a}_{\bm{k}}, \hat{a}^\dag_{\bm{k}'}]=\delta_{\bm{k},\bm{k}'}$, $[\hat{b}_{\bm{k}}, \hat{b}^\dag_{\bm{k}'}]=\delta_{\bm{k},\bm{k}'}$, $[\hat{a}_{\bm{k}}, \hat{a}_{\bm{k}'}]=0$, and $[\hat{b}_{\bm{k}}, \hat{b}_{\bm{k}'}]=0$, it follows that
\begin{align}
S^+_{i\in A}&=\sqrt{2S/N}\sum_{\bm{k}}e^{-i\bm{k}\cdot\bm{r}_i}\hat{a}_{\bm{k}},\nonumber\\
S^+_{i\in B}&=\sqrt{2S/N}\sum_{\bm{k}}e^{-i\bm{k}\cdot\bm{r}_i}\hat{b}^\dag_{\bm{k}},\nonumber\\
S^z_{i\in A}&=S-(1/N)\sum_{\bm{k},\bm{k}'}e^{-i(\bm{k}-\bm{k}')\cdot\bm{r}_i}\hat{a}_{\bm{k}}^\dag \hat{a}_{\bm{k}'},\nonumber\\
S^z_{i\in B}&=-S+(1/N)\sum_{\bm{k},\bm{k}'}e^{-i(\bm{k}-\bm{k}')\cdot\bm{r}_i}\hat{b}_{\bm{k}}^\dag \hat{b}_{\bm{k}'},
\end{align}
where $S$ is the spin number, $N$ is the number of lattice sites in each sublattice, and $S^\pm_i=S_i^x\pm iS_i^y$.
In order to diagonalize the Hamiltonian~(\ref{Original-Hamiltonian-AFM}) in terms of magnon operators, we next perform a generalized Bogoliubov transformation \cite{Kamra2017},
\begin{align}
\begin{bmatrix}
\hat{m}_{\alpha, \bm{k}}\\
\hat{m}_{\beta,-\bm{k}}^\dag\\
\hat{m}_{\alpha,-\bm{k}}^\dag\\
\hat{m}_{\beta,\bm{k}}
\end{bmatrix}
=
\begin{bmatrix}
u_{\alpha,a} & v_{\alpha,b} & v_{\alpha,a} & u_{\alpha,b} \\
v^*_{\beta,a} & u^*_{\beta,b} & u^*_{\beta,a} & v^*_{\beta,b} \\
v^*_{\alpha,a} & u^*_{\alpha,b} & u^*_{\alpha,a} & v^*_{\alpha,b} \\
u_{\beta,a} & v_{\beta,b} & v_{\beta,a} & u_{\beta,b}
\end{bmatrix}
\begin{bmatrix}
\hat{a}_{\bm{k}}\\
\hat{b}^\dag_{-\bm{k}}\\
\hat{a}^\dag_{-\bm{k}}\\
\hat{b}_{\bm{k}}
\end{bmatrix},
\label{Bogoliubov-transformation}
\end{align}
or equivalently,
\begin{align}
\begin{bmatrix}
\hat{a}_{\bm{k}}\\
\hat{b}^\dag_{-\bm{k}}\\
\hat{a}^\dag_{-\bm{k}}\\
\hat{b}_{\bm{k}}
\end{bmatrix}
=
\begin{bmatrix}
u_{a,\alpha} & v_{a,\beta} & v_{a,\alpha} & u_{a,\beta} \\
v^*_{b,\alpha} & u^*_{b,\beta} & u^*_{b,\alpha} & v^*_{b,\beta} \\
v^*_{a,\alpha} & u^*_{a,\beta} & u^*_{a,\alpha} & v^*_{a,\beta} \\
u_{b,\alpha} & v_{b,\beta} & v_{b,\alpha} & u_{b,\beta}
\end{bmatrix}
\begin{bmatrix}
\hat{m}_{\alpha, \bm{k}}\\
\hat{m}_{\beta,-\bm{k}}^\dag\\
\hat{m}_{\alpha,-\bm{k}}^\dag\\
\hat{m}_{\beta,\bm{k}}
\end{bmatrix},
\label{Bogoliubov-transformation2}
\end{align}
where the matrix elements in Eqs.~(\ref{Bogoliubov-transformation}) and (\ref{Bogoliubov-transformation2}) are determined under the condition such that the operators $\hat{m}_{\alpha,\bm{k}}$ and $\hat{m}_{\beta,\bm{k}}$ must satisfy the relations $[\hat{m}_{\alpha,\bm{k}}, H_{\mathrm{AFM}}]=\hbar\omega_{\alpha,\bm{k}}\hat{m}_{\alpha,\bm{k}}$ and $[\hat{m}_{\beta,\bm{k}}, H_{\mathrm{AFM}}]=\hbar\omega_{\beta,\bm{k}}\hat{m}_{\beta,\bm{k}}$.

Finally, the original Hamiltonian~(\ref{Original-Hamiltonian-AFM}) is expressed in a diagonal form in terms of the creation and annihilation operators of antiferromagnetic magnons as
\begin{align}
H_{\mathrm{AFM}}=\sum_{\bm{k}}\left[\hbar\omega_{\alpha,\bm{k}}\hat{m}_{\alpha,\bm{k}}^\dag\hat{m}_{\alpha,\bm{k}}+\hbar\omega_{\beta,\bm{k}}\hat{m}_{\beta,\bm{k}}^\dag\hat{m}_{\beta,\bm{k}}\right],
\end{align}
where $\alpha$ and $\beta$ denote the two antiferromagnetic magnon modes and $\bm{k}$ is the wave vector of the magnons.
Henceforth we focus on the antiferromagnetic resonance state where all the spins are precessing uniformly (i.e., with $\bm{k}=0$) and drop the wave-vector dependence in $\omega_{\mu,\bm{k}}$ and $\hat{m}_{\mu,\bm{k}}$.
The antiferromagnetic resonance frequencies are given by $\omega_{\alpha,\beta}^2=\omega_E(2\omega_\parallel+\omega_\perp)+\omega_H^2\pm\sqrt{\omega_E^2\omega_\perp^2+4\omega_H^2\omega_E(2\omega_\parallel+\omega_\perp)}$ with $\omega_\alpha\ge\omega_\beta$ \cite{Kamra2017,Johansen2018}.
Here, the frequencies are given by $\omega_E= SZJ/\hbar$, $\omega_H=|\gamma|B_0$, $\omega_\parallel=S K_\parallel/\hbar$, and $\omega_\perp=S K_\perp/\hbar$, with $Z$ being the number of nearest neighbors in the antiferromagnet.
A schematic illustration of the magnetic-field dependence of the resonance frequencies $\omega_{\alpha}$ and $\omega_\beta$ of easy-axis antiferromagnets is shown in Fig.~\ref{Fig1}(b).

\section{Microwave-Magnon Interaction in Antiferromagnets \label{Sec-Microwave-Magnon}}
In this section, following Refs.~\cite{Johansen2018,Boventer2023}, we consider the interaction between the antiferromagnetic magnons and microwave cavity photons.
Generically, the interaction between the magnetization $\bm{M}$ and a magnetic field $\bm{B}$ comes from the Zeeman interaction $\bm{M}\cdot\bm{B}$.
In the antiferromagnetic resonance state, the spins are precessing around the ground-state direction (the $z$ axis in our case).
Thus, the small fluctuation of the magnetization density around the ground-state direction $\delta\bm{m}_\perp=\delta m_x\bm{e}_x+\delta m_y\bm{e}_y$, which is equivalent to the magnon excitation, couples with the applied ac magnetic field.
Here, $\delta m_x=S^x_{i\in A}+S^x_{i\in B}$ and $\delta m_y=S^y_{i\in A}+S^y_{i\in B}$.

For concreteness, let us consider the case of easy-axis antiferromagnets with $K_\perp=0$.
In this case, we have \cite{Rezende2019}
\begin{align}
&u_{a,\alpha}=u_{b,\beta}\equiv U=\frac{1}{\sqrt{2}}\sqrt{\frac{\omega_E+\omega_\parallel}{\sqrt{\omega_\parallel\left(2\omega_E+\omega_\parallel\right)}}+1},\nonumber\\
&v_{a,\beta}=v_{b,\alpha}\equiv V=-\frac{1}{\sqrt{2}}\sqrt{\frac{\omega_E+\omega_\parallel}{\sqrt{\omega_\parallel\left(2\omega_E+\omega_\parallel\right)}}-1},
\end{align}
and the other coefficients are zero.
Then, the explicit forms of the magnetization density components $\delta m_x$ and $\delta m_y$ in the antiferromagnetic resonance state is given by
\begin{align}
\delta m_x&=\sqrt{S/2N}(U+V)\left(\hat{m}_\alpha+\hat{m}_\alpha^\dag+\hat{m}_\beta+\hat{m}_\beta^\dag\right),\nonumber\\
\delta m_y&=(1/i)\sqrt{S/2N}(U+V)\left(\hat{m}_\alpha-\hat{m}_\alpha^\dag-\hat{m}_\beta+\hat{m}_\beta^\dag\right)
\end{align}
In general, the strength of the easy-axis anisotropy is much smaller than that of the exchange interaction, i.e., $\omega_\parallel \ll \omega_E$.
In this limit, we get $U+V\approx (\omega_\parallel / 2\omega_E)^{1/4}$ to leading order in $\omega_\parallel / \omega_E$.

The coupling between microwave photons and antiferromagnetic magnons originates from the Zeeman interaction term in Eq.~(\ref{Original-Hamiltonian-AFM}): $H_{\mathrm{Zeeman}}=|\gamma|N \delta\bm{m}_\perp\cdot\bm{B}$.
To see this, we consider the quantized magnetic field in a microwave cavity of the form
\begin{align}
\bm{B}(\bm{r})=\sum_{n,\lambda}i\cos\left(\frac{n\pi r_j}{L}\right)\sqrt{\frac{\hbar \omega_n \mu_0}{V_{\mathrm{c}}}}\left(\hat{a}_{n\lambda}\bm{e}_j\times\bm{e}_\lambda-\hat{a}^\dag_{n\lambda}\bm{e}_j\times\bm{e}_\lambda^*\right),
\end{align}
where we have assumed that the magnetic field is propagating only in the $j$ direction.
Here, $\mu_0$ is the vacuum permeability, $L$ is the length of the cavity in the $j$ direction, $V_{\mathrm{c}}$ is the volume of the cavity, $n$ is the frequency mode of the cavity photon, $\lambda$ denotes the polarization, $\bm{e}_\lambda$ is the polarization unit vector, and $\hat{a}_{n\lambda}$ is the annihilation operator of a photon with mode $n$ and polarization $\lambda$.

We assume a magnetic field propagating in the $y$ direction with a circular polarization basis $\bm{e}_\pm=(\bm{e}_x\mp i\bm{e}_z)/\sqrt{2}$.
Also, we consider only the lowest-energy cavity mode $n=1$ with $\omega_1\equiv \omega_{\mathrm{e}}$.
Then, the resultant Hamiltonian describing the coupling between microwave photons and antiferromagnetic magnons reads
\begin{align}
H_g=\sum_{\lambda=\pm}\sum_{\mu=\alpha,\beta}\hbar\lambda g_\mu\left(\hat{a}_\lambda^\dag \hat{m}_\mu+\hat{m}^\dag_\mu \hat{a}_\lambda\right),
\label{Hamiltonian-H_g}
\end{align}
where
\begin{align}
g_\alpha=g_\beta=g_0\left(\frac{\omega_\parallel}{8\omega_E}\right)^{1/4}\sqrt{2SN}
\label{microwave-magnon-coupling-strength}
\end{align}
with $g_0=\eta|\gamma|\sqrt{\hbar \omega_{\mathrm{e}}\mu_0/(4V_{\mathrm{c}})}$.
Here, $\eta\le 1$ is the spatial overlap factor between the cavity mode and the antiferromagnet and $V_{\mathrm{c}}$ is the volume of the cavity.
We note that only the expression for the coupling strength for the lower-energy mode $g_\beta$ in Eq.~(\ref{microwave-magnon-coupling-strength}) was obtained in Ref.~\cite{Johansen2018}.
In what follows, we drop the $\lambda$ dependence in Eq.~(\ref{Hamiltonian-H_g}) for simplicity.

\section{Light-Magnon Interaction in Antiferromagnets \label{Sec-Light-Magnon}}
In this section, we derive the light-magnon interaction in antiferromagnets in the cases with and without an optical cavity, both from a linear magneto-optic (Faraday) effect in antiferromagnets.
See Appendix~\ref{Appendix-Derivation} for a brief derivation and for a comparison with the the light-magnon interaction in ferromagnets.
For concreteness, as in Sec.~\ref{Sec-Microwave-Magnon}, we focus on the easy-axis antiferromagnets with $K_\perp=0$.

\subsection{The case with an optical cavity \label{Sec-Light-magnon-with-optical-cavity}}
First, let us consider the case with an optical cavity.
It has been proposed that the interaction between a circularly polarized light propagating in the direction perpendicular to the $z$ axis and the antiferromagnetic magnons is described by \cite{Parvini2020}
\begin{align}
H_\zeta=-\hbar\hat{b}^\dag\hat{b}\left[G_\alpha\left(\hat{m}_\alpha^\dag+\hat{m}_\alpha\right)+G_\beta\left(\hat{m}_\beta^\dag+\hat{m}_\beta\right)\right],
\label{Hamiltonian-Parvini2020}
\end{align}
where $\hat{b}$ is the annihilation operator of a photon with a (right or left) circular polarization.
If we assume an equal (absolute) value of the Faraday rotation angles for the two sublattices $A$ and $B$, the coupling strength $G_\mu$ ($\mu=\alpha, \beta$) is given by \cite{Parvini2020}
\begin{align}
G_\mu=\frac{c\theta_{\mathrm{F}}}{4\sqrt{\varepsilon_{\mathrm{r}}}}\frac{\kappa_\mu}{\sqrt{2SN}},
\label{light-magnon-coupling-strength-G}
\end{align}
where $\kappa_\mu$ is determined from the coefficients in the Bogoliubov transformation~(\ref{Bogoliubov-transformation}), $c$ is the speed of light, $\theta_{\mathrm{F}}$ is the Faraday rotation angle of each sublattice per unit length, and $\varepsilon_{\mathrm{r}}$ is the relative permittivity of the antiferromagnet.
In the case of easy-axis antiferromagnets where the condition $\omega_\parallel / \omega_E\ll 1$ is usually satisfied, we have
\begin{align}
\kappa_{\alpha,\beta}=\left(\frac{\omega_\parallel}{2\omega_E}\right)^{1/4}\pm K\left(\frac{2\omega_E}{\omega_\parallel}\right)^{1/4},
\label{light-magnon-coupling-strength-kappa}
\end{align}
where $K=K_-/K_+$ denotes an intrinsic magneto-optic asymmetry between two sublattices (see Appendix~\ref{Appendix-Derivation}) \cite{Parvini2020}.
Note that the expression for $\kappa_\mu$ in Eq.~(\ref{light-magnon-coupling-strength-kappa}) is independent of the external magnetic field.

As is well known, the presence of a cavity enhances the coupling strength.
To see this, we define the steady-state cavity population as $\bar{n}_{\mathrm{cav}}=\langle\hat{b}^\dag\hat{b}\rangle$.
The photon operator can be split into an average coherent amplitude $\langle \hat{b}\rangle$ and a fluctuating term $\delta\hat{b}$ as $\hat{b}=\langle\hat{b}\rangle+\delta\hat{b}$ \cite{Aspelmeyer2014}.
Substituting this expression into Eq.~(\ref{Hamiltonian-Parvini2020}), we obtain
\begin{align}
H_\zeta=-\hbar\left[\zeta_\alpha\left(\hat{m}_\alpha^\dag+\hat{m}_\alpha\right)+\zeta_\beta\left(\hat{m}_\beta^\dag+\hat{m}_\beta\right)\right]\left(\delta\hat{b}^\dag+\delta\hat{b}\right),
\label{Light-magnon-interaction-with-cavity}
\end{align}
where we have assumed without loss of generality that $\langle\hat{b}\rangle$ is real valued as $\langle\hat{b}\rangle=\sqrt{\bar{n}_{\mathrm{cav}}}$ and we have retained only the terms linear in $\delta\hat{b}$.
The coupling strength $\zeta_\mu$ ($\mu=\alpha, \beta$) is given by
\begin{align}
\zeta_\mu=G_\mu\sqrt{\bar{n}_{\mathrm{cav}}}.
\label{zeta-with-cavity}
\end{align}
Typically, the steady-state photon number in an optical cavity can be $\bar{n}_{\mathrm{cav}}\sim 10^6$ \cite{Zhu2020}.
In what follows, we replace $\delta\hat{b}$ by $\hat{b}$ in Eq.~(\ref{Light-magnon-interaction-with-cavity}) for simplicity of notation.

In the present context, the meaning of the Faraday effect in antiferromagnets may be unclear.
Generically, the antiferromagnetic magnons can couple to photons through one-magnon and two-magnon Raman scattering processes.
It is known that in antiferromagnets the magnitude of the one-magnon Raman scattering is rather small compared to that of the two-magnon Raman scattering \cite{Fleury1968,Lockwood1987}.
However, we here would like to emphasize that the light-magnon interaction in our study, which corresponds to a one-magnon scattering process, can be significantly enhanced by the presence of an optical cavity with the factor $\sqrt{\bar{n}_{\mathrm{cav}}}\sim 10^3$ [see Eq.~(\ref{zeta-with-cavity})].
This one-magnon scattering, characterized by the factor $\theta_{\mathrm{F}}$, describes the Faraday rotation per sublattice \cite{Parvini2020} (see also Appendix~\ref{Appendix-Derivation}), which means that the total Faraday rotation angle can be zero at zero magnetic field in simple two-sublattice systems as is intuitively understood.
On the one hand, it has also been suggested that in some antiferromagnets under magnetic fields the linear magneto-optic coupling (i.e., the Faraday rotation angle) can be as large as that of the ferromagnet YIG \cite{Bi2008}.
Due to the lack of the experimental data on the value of the Faraday rotation angle per sublattice in antiferromagnets, in Sec.~\ref{Sec-Numerical-Results} we shall firstly compute the magnitude of the transduction efficiency as a function of $\theta_{\mathrm{F}}$.
After that,  for concreteness we shall adopt the value for YIG as an estimate of $\theta_{\mathrm{F}}$.

\subsection{The case without an optical cavity \label{Sec-Light-magnon-wo-optical-cavity}}
Next, let us consider the case without an optical cavity.
In this case, we also start with Eq.~(\ref{Hamiltonian-Parvini2020}) but consider it in a generalized form  \cite{Parvini2020}:
\begin{align}
H_\zeta=-\hbar \left(\hat{b}_R^\dag\hat{b}_R-\hat{b}_L^\dag\hat{b}_L\right)\left[G_\alpha\left(\hat{m}_\alpha^\dag+\hat{m}_\alpha\right)+G_\beta\left(\hat{m}_\beta^\dag+\hat{m}_\beta\right)\right],
\label{Light-magnon-interaction-without-cavity-1}
\end{align}
where $\hat{b}_{R,L}$ is the annihilation operator of a photon with a right-circular (left-circular) polarization.
For concreteness, let us consider an incident light propagating in the $x$ direction (which is a direction perpendicular to the ground-state direction of the magnetic moments).
For a strong $z$-polarized light we have $\hat{b}_{R,L}(t)=(\hat{b}_y\mp i\hat{b}_z)/\sqrt{2}\simeq (\hat{b}_y\mp i\langle\hat{b}_z\rangle)/\sqrt{2}$, where $\langle\hat{b}_z\rangle=\sqrt{P_0/(\hbar\Omega_0)}e^{-i\Omega_0 t}$ with $P_0$ ($\Omega_0$) the power (angular frequency) of the incident light \cite{Hammerer2010}.
Because in the present case $\hat{b}$ and $\hat{b}^\dag$ represent itinerant photon operators whose dimension is $[\mathrm{time}]^{-1/2}$, the interaction Hamiltonian~(\ref{Light-magnon-interaction-without-cavity-1}) needs to be integrated over the interaction time \cite{Sekine2024}.

In order for the interaction time $\tau=d_{\mathrm{AFI}}/c$ with $d_{\mathrm{AFI}}$ the thickness of the antiferromagnet to be much shorter than the time scale of the antiferromagnetic dynamics (i.e., the antiferromagnetic resonance frequency) $1/\omega_\mu\sim 10^{-12}\, \mathrm{s}$, the thickness $d_{\mathrm{AFI}}$ needs to be of, or shorter than, $\mathcal{O}(1\, \mathrm{\mu m})$.
When this condition is satisfied, the integrand, i.e., the operators in Eq.~(\ref{Light-magnon-interaction-without-cavity-1}) can be regarded as constant during the interaction, enabling the integration over the interaction time as $\int_0^{\tau}dt=\tau$.
Then, setting $\hat{b}_y\equiv \hat{b}_{\mathrm{in}}$, we arrive at the interaction Hamiltonian of the form
\begin{align}
H_\xi=&-i\hbar\left[\sqrt{\xi_\alpha}\left(\hat{m}_\alpha^\dag+\hat{m}_\alpha\right)+\sqrt{\xi_\beta}\left(\hat{m}_\beta^\dag+\hat{m}_\beta\right)\right]\nonumber\\
&\times\left(\hat{b}_{\mathrm{in}}e^{i\Omega_0 t}-\hat{b}^\dag_{\mathrm{in}}e^{-i\Omega_0 t}\right),
\label{Light-magnon-interaction-without-cavity-2}
\end{align}
which is indeed the sum of the beam-splitter type and parametric-amplification type interactions representing the Faraday effect \cite{Hisatomi2016,Hammerer2010}.
The light-magnon interaction strength $\xi_\mu$ ($\mu=\alpha,\beta$) is given by
\begin{align}
\xi_\mu=G_\mu^2\frac{d_{\mathrm{AFI}}^2}{c^2}\frac{P_0}{\hbar\Omega_0}=\frac{\kappa_\mu^2}{32\varepsilon}\frac{\phi_{\mathrm{F}}^2}{SN}\frac{P_0}{\hbar\Omega_0}
\label{Light-magnon-interaction-without-cavity-strength}
\end{align}
with $\phi_{\mathrm{F}}=\theta_{\mathrm{F}}d_{\mathrm{AFI}}$ being the total Faraday rotation angle per sublattice of the antiferromagnet (which should not be confused with $\theta_{\mathrm{F}}$).

\section{Microwave-to-Optical Quantum Transduction \label{Sec-Quantum-Transduction}}
Having the key ingredients obtained so far, i.e., the strengths of the microwave-magnon interaction and the light-magnon interaction, we are now in a position to derive the explicit expressions for the transduction efficiency between the microwave range and the optical range in antiferromagnets.
The Hamiltonian of the system is given by
\begin{align}
H_{\mathrm{sys}}=H_0+H_g+H_{\zeta(\xi)},
\end{align}
where $H_0=\hbar \omega_{\mathrm{e}}\hat{a}^\dag\hat{a}-\hbar\delta \omega_{\mathrm{o}}\hat{b}^\dag\hat{b}+\hbar\omega_\alpha\hat{m}_\alpha^\dag\hat{m}_\alpha+\hbar\omega_\beta\hat{m}_\beta^\dag\hat{m}_\beta$ for systems with microwave and optical cavities, while $H_0=\hbar \omega_{\mathrm{e}}\hat{a}^\dag\hat{a}+\hbar\omega_\alpha\hat{m}_\alpha^\dag\hat{m}_\alpha+\hbar\omega_\beta\hat{m}_\beta^\dag\hat{m}_\beta$ for systems with a microwave cavity but without an optical cavity.
Here, $\delta \omega_{\mathrm{o}}\equiv \omega_{\mathrm{P}}-\omega_{\mathrm{o}}$ is the detuning of the optical cavity frequency from the pump frequency \cite{Zhu2020}.

\begin{figure*}[!t]
\centering
\includegraphics[width=1.6\columnwidth]{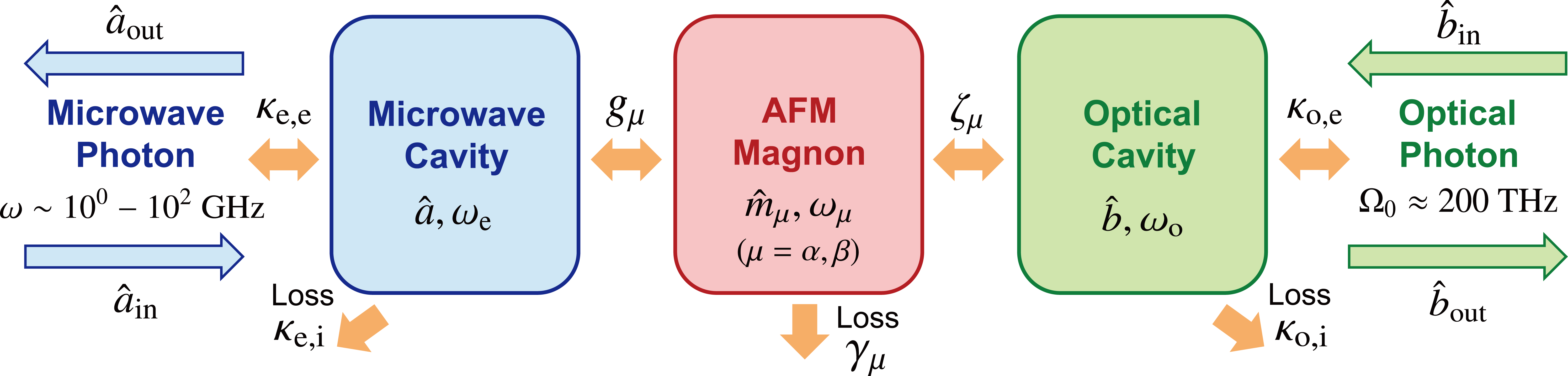}
\caption{Schematic illustration of our setup for the quantum transduction with microwave and optical cavities in terms of the operators ($\hat{a}_{\mathrm{in}}$, $\hat{a}_{\mathrm{out}}$, $\hat{a}$, $\hat{m}_\mu$, $\hat{b}$, $\hat{b}_{\mathrm{in}}$, and $\hat{b}_{\mathrm{out}}$), frequencies ($\omega$, $\omega_{\mathrm{e}}$, $\omega_\mu$, $\omega_{\mathrm{o}}$, and $\Omega_0$), coupling strengths ($\kappa_{\mathrm{e,e}}$, $g_\mu$, $\zeta_\mu$, and $\kappa_{\mathrm{o,e}}$), and losses ($\kappa_{\mathrm{e,i}}$, $\gamma_\mu$, and $\kappa_{\mathrm{o,i}}$).
}\label{Fig2}
\end{figure*}
\subsection{General consideration}
We consider a generic description to obtain an expression for the transduction efficiency \cite{Han2021}, which can be applied to the setups with and without an optical cavity.
Let us define the vectors $\vec{c}=[\hat{a},\hat{m}_\alpha,\hat{m}_\beta,\hat{b}]^T$, $\vec{c}_{\mathrm{in}}=[\hat{a}_{\mathrm{in}},0,0,\hat{b}_{\mathrm{in}}]^T$, and $\vec{c}_{\mathrm{out}}=[\hat{a}_{\mathrm{out}},0,0,\hat{b}_{\mathrm{out}}]^T$.
Here, $\hat{a}_{\mathrm{in}}$ and $\hat{a}_{\mathrm{out}}$ ($\hat{b}_{\mathrm{in}}$ and $\hat{b}_{\mathrm{out}}$) are the incoming and outgoing itinerant microwave (optical) photons, respectively.
Then, the equations of motion $\dot{\hat{c}}_j=(i/\hbar)[H_{\mathrm{total}},\hat{c}_j]$, with $\hat{c}_j$ being the $j$-th component of $\hat{c}$ and $H_{\mathrm{total}}$ being the total Hamiltonian of the system including the bath Hamiltonian, in the matrix representation can be written in a generic form
\begin{align}
\dot{\vec{c}}=-A\vec{c}-B\vec{c}_{\mathrm{in}},
\label{equation-of-motion-generic-form}
\end{align}
where $A$ and $B$ are $4\times 4$ matrices.
We employ the standard input-output formalism to take into account the presence of the incoming and outgoing itinerant photons, giving rise to
\begin{align}
\vec{c}_{\mathrm{out}}=\vec{c}_{\mathrm{in}}+B^T\vec{c},
\label{input-output-generic-form}
\end{align}
where $T$ is the transpose of a matrix.
Combining Eqs.~(\ref{equation-of-motion-generic-form}) and (\ref{input-output-generic-form}), the scattering matrix $S$ that connects the incoming and outgoing itinerant photons is introduced to obtain
\begin{align}
\vec{c}_{\mathrm{out}}=S\vec{c}_{\mathrm{in}},
\end{align}
where $S=I_{4\times 4}-B^T[-i\omega I_{4\times 4}+A]^{-1}B$ with $I_{4\times 4}$ being the $4\times 4$ identity matrix and we have used the Fourier transform defined by $\vec{c}(t)=\int dt/(2\pi)\, e^{-i\omega t}\vec{c}(\omega)$.
Finally, the microwave-to-optical transduction efficiency $\eta$ is defined by 
\begin{align}
\eta=\left|\frac{\langle\hat{b}_{\mathrm{out}}\rangle}{\langle\hat{a}_{\mathrm{in}}\rangle}\right|^2=|S_{4,1}|^2=|S_{1,4}|^2,
\label{efficiency-definition}
\end{align}
with $S_{i,j}$ being an matrix element of $S$.
In a similar way as the transduction efficiency $\eta$, we can obtain other quantities from the matrix elements $S_{i,j}$, such as the reflection coefficient for the itinerant microwave mode $|S_{1,1}|^2$.

\subsection{The case with an optical cavity}
First, let us consider the microwave-to-optical quantum transduction in the case with an optical cavity, as schematically depicted in Fig.~\ref{Fig2}.
The equations of motion including losses for the microwave cavity photon $\hat{a}$, the antiferromagnetic magnons $\hat{m}_\mu$ ($\mu=\alpha,\beta$), and the optical cavity photon $\hat{b}$ are, respectively,
\begin{align}
\dot{\hat{a}}
=-i\omega_{\mathrm{e}}\hat{a}-i\sum_\mu g_\mu\hat{m}_\mu-\frac{\kappa_{\mathrm{e}}}{2}\hat{a}-\sqrt{\kappa_{\mathrm{e,e}}}\hat{a}_{\mathrm{in}},
\end{align}
\begin{align}
\dot{\hat{m}}_\mu
=-i\omega_{\mu}\hat{m}_\mu-ig_\mu\hat{a}-i\zeta_\mu\hat{b}-\frac{\gamma_\mu}{2}\hat{m}_\mu,
\end{align}
and
\begin{align}
\dot{\hat{b}}
=i\delta\omega_{\mathrm{o}}\hat{b}-i\sum_\mu\zeta_\mu\hat{m}_\mu-\frac{\kappa_{\mathrm{o}}}{2}\hat{b}-\sqrt{\kappa_{\mathrm{o,e}}}\hat{b}_{\mathrm{in}},
\end{align}
where $\kappa_{\mathrm{e}}=\kappa_{\mathrm{e,e}}+\kappa_{\mathrm{e,i}}$ and $\kappa_{\mathrm{o}}=\kappa_{\mathrm{o,e}}+\kappa_{\mathrm{o,i}}$.
The incoming and outgoing itinerant photons are related by the input-output formalism as
\begin{align}
\hat{a}_{\mathrm{out}}&=\hat{a}_{\mathrm{in}}+\sqrt{\kappa_{\mathrm{e,e}}}\hat{a},\nonumber\\
\hat{b}_{\mathrm{out}}&=\hat{b}_{\mathrm{in}}+\sqrt{\kappa_{\mathrm{o,e}}}\hat{b}.
\end{align}

Combining the above ingredients, the matrices $A$ and $B$ in Eq.~(\ref{equation-of-motion-generic-form}) are obtained as
\begin{align}
A&=
\begin{bmatrix}
i\omega_{\mathrm{e}}+\kappa_{\mathrm{e}}/2 & ig_\alpha & ig_\beta & 0\\
ig_\alpha & i\omega_\alpha+\gamma_\alpha/2 & 0 & i\zeta_\alpha\\
ig_\beta & 0 & i\omega_\beta+\gamma_\beta/2 & i\zeta_\beta\\
0 & i\zeta_\alpha & i\zeta_\beta & -i\delta\omega_{\mathrm{o}}+\kappa_{\mathrm{o}}/2
\end{bmatrix}
\end{align}
and
\begin{align}
B&=
\begin{bmatrix}
\sqrt{\kappa_{\mathrm{e,e}}} & 0 & 0 & 0\\
0 & 0 & 0 & 0\\
0 & 0 & 0 & 0\\
0 & 0 & 0 & \sqrt{\kappa_{\mathrm{o,e}}}
\end{bmatrix},
\end{align}
respectively.
Finally, introducing the susceptibilities $\chi_{\mathrm{e}}=[-i(\omega-\omega_{\mathrm{e}})+\kappa_{\mathrm{e}}/2]^{-1}$, $\chi_\mu=[-i(\omega-\omega_\mu)+\gamma_\mu/2]^{-1}$, and $\chi_{\mathrm{o}}=[-i(\omega+\delta\omega_{\mathrm{o}})+\kappa_{\mathrm{o}}/2]^{-1}$, the transduction efficiency $\eta$ of the microwave-to-optical quantum transduction is obtained from Eq.~(\ref{efficiency-definition}) as
\begin{widetext}
\begin{align}
\eta=\left|
\frac{\sqrt{\kappa _{\mathrm{e,e}}} \sqrt{\kappa _{\mathrm{o,e}}} \left(\zeta _\beta g_\beta\chi_\beta+\zeta _\alpha g_\alpha\chi_\alpha\right)}{\zeta _\beta^2\chi_\beta\chi _{\mathrm{e}}^{-1}+\zeta _\alpha^2\chi_\alpha\chi_{\mathrm{e}}^{-1}+\chi_{\mathrm{e}}^{-1}\chi_{\mathrm{o}}^{-1}+\left(\zeta _\alpha^2 g_\beta^2+\zeta _\beta^2 g_\alpha^2-2 \zeta _\alpha \zeta _\beta g_\alpha g_\beta\right)\chi_\alpha\chi_\beta+g_\beta^2\chi_\beta\chi_{\mathrm{o}}^{-1}+g_\alpha^2\chi_\alpha\chi_{\mathrm{o}}^{-1}}
\right|^2.
\label{transduction-efficiency-w-optical-cavity1}
\end{align}
\end{widetext}
When only one of the two antiferromagnetic magnon modes ($\alpha\ \mathrm{or}\ \beta$) is excited, the transduction efficiency $\eta$ reduces to a simpler expression,
\begin{align}
\eta=\left|
\frac{\sqrt{\kappa _{\mathrm{e,e}}} \sqrt{\kappa _{\mathrm{o,e}}}\zeta _\mu g_\mu}{\zeta_\mu^2\chi_{\mathrm{e}}^{-1}+g_\mu^2\chi_{\mathrm{o}}^{-1}+\chi_{\mu}^{-1} \chi_{\mathrm{e}}^{-1} \chi_{\mathrm{o}}^{-1}}
\right|^2
\label{transduction-efficiency-w-optical-cavity2}
\end{align}
with $\mu=\alpha$ or $\beta$, which takes the same form as in the case of the microwave-to-optical quantum transduction with ferromagnets \cite{Zhu2020,Han2021}.

\subsection{The case without an optical cavity}
\begin{figure}[!b]
\centering
\includegraphics[width=\columnwidth]{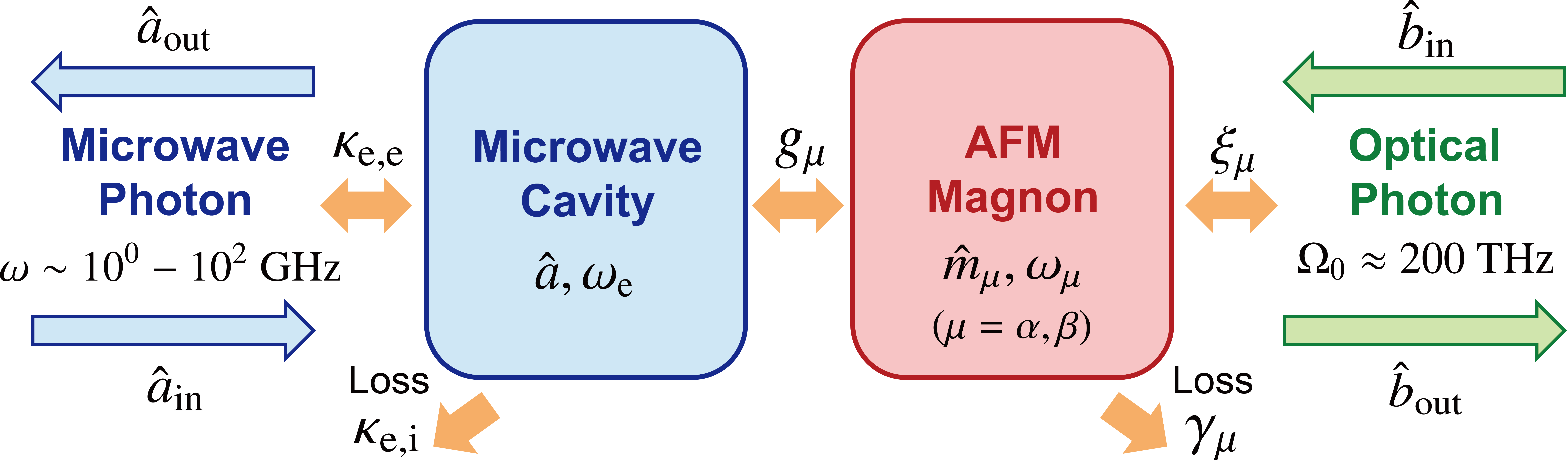}
\caption{Schematic illustration of our setup for the quantum transduction with a microwave cavity but without an optical cavity in terms of the operators ($\hat{a}_{\mathrm{in}}$, $\hat{a}_{\mathrm{out}}$, $\hat{a}$, $\hat{m}_\mu$, $\hat{b}_{\mathrm{in}}$, and $\hat{b}_{\mathrm{out}}$), frequencies ($\omega$, $\omega_{\mathrm{e}}$, $\omega_\mu$, and $\Omega_0$), coupling strengths ($\kappa_{\mathrm{e,e}}$, $g_\mu$, and $\xi_\mu$), and losses ($\kappa_{\mathrm{e,i}}$ and $\gamma_\mu$).
}\label{Fig3}
\end{figure}

Next, let us consider the microwave-to-optical quantum transduction in the case without an optical cavity, as schematically depicted in Fig.~\ref{Fig3}.
The equations of motion including losses for the microwave cavity photon $\hat{a}$ and the antiferromagnetic magnons $\hat{m}_\mu$ ($\mu=\alpha,\beta$) are, respectively,
\footnote{The factor $e^{\pm i\Omega_0 t}$ in Eq.~(\ref{Light-magnon-interaction-without-cavity-2}) can be eliminated when performing the Fourier transform as follows:
$e^{i\Omega_0 t}\hat{b}_{\mathrm{in}}(t)=\int (d\omega/2\pi)\, \hat{b}_{\mathrm{in}}(\omega)e^{-i(\omega-\Omega_0)t}$,
$\hat{m}_\mu(t)=\int (d\omega/2\pi)\, \hat{m}_\mu(\omega)e^{-i\omega t}=\int (d\omega/2\pi)\, \hat{m}_\mu(\omega-\Omega_0)e^{-i(\omega-\Omega_0)t}$, and
$\hat{a}(t)=\int (d\omega/2\pi)\, \hat{a}(\omega)e^{-i\omega t}=\int (d\omega/2\pi)\, \hat{a}(\omega-\Omega_0)e^{-i(\omega-\Omega_0)t}$.
}
\begin{align}
\dot{\hat{a}}
=-i\omega_{\mathrm{e}}\hat{a}-i\sum_\mu g_\mu\hat{m}_\mu-\frac{\kappa_{\mathrm{e}}}{2}\hat{a}-\sqrt{\kappa_{\mathrm{e,e}}}\hat{a}_{\mathrm{in}},
\end{align}
and
\begin{align}
\dot{\hat{m}}_\mu
=-i\omega_{\mu}\hat{m}_\mu-ig_\mu\hat{a}-\frac{\gamma_\mu}{2}\hat{m}_\mu-\sqrt{\xi_\mu}\hat{b}_{\mathrm{in}},
\end{align}
where $\kappa_{\mathrm{e}}=\kappa_{\mathrm{e,e}}+\kappa_{\mathrm{e,i}}$.
The incoming and outgoing itinerant photons are related by the input-output formalism as
\begin{align}
\hat{a}_{\mathrm{out}}&=\hat{a}_{\mathrm{in}}+\sqrt{\kappa_{\mathrm{e,e}}}\hat{a},\nonumber\\
\hat{b}_{\mathrm{out}}&=\hat{b}_{\mathrm{in}}+\sqrt{\xi_\alpha}\hat{m}_\alpha+\sqrt{\xi_\beta}\hat{m}_\beta.
\end{align}

Combining the above ingredients, the matrices $A$ and $B$ in Eq.~(\ref{equation-of-motion-generic-form}) are obtained as
\footnote{In this case, we need to put a dummy constant $\Delta$ in the matrix element $A_{4,4}$ of Eq.~(\ref{Matrix-element-A-without-cavity}) to obtain the inverse matrix of $A$. The final expression for $\eta$ does not depend on $\Delta$.}
\begin{align}
A&=
\begin{bmatrix}
i\omega_{\mathrm{e}}+\kappa_{\mathrm{e}}/2 & ig_\alpha & ig_\beta & 0\\
ig_\alpha &i\omega_\alpha+\gamma_\alpha/2 & 0 & 0\\
ig_\beta & 0 & i\omega_\beta+\gamma_\beta/2 & 0\\
0 & 0 & 0 & 0
\end{bmatrix}
\label{Matrix-element-A-without-cavity}
\end{align}
and
\begin{align}
B&=
\begin{bmatrix}
\sqrt{\kappa_{\mathrm{e,e}}} & 0 & 0 & 0\\
0 & 0 & 0 & \sqrt{\xi_\alpha}\\
0 & 0 & 0 & \sqrt{\xi_\beta}\\
0 & 0 & 0 & 0
\end{bmatrix},
\end{align}
respectively.
Finally, the transduction efficiency $\eta$ of the microwave-to-optical quantum transduction is obtained from Eq.~(\ref{efficiency-definition}) as
\begin{align}
\eta=\left|
\frac{\sqrt{\kappa_{\mathrm{e,e}}} \left(\sqrt{\xi_\alpha} g_\alpha \chi_{\alpha}+\sqrt{\xi_\beta} g_\beta \chi_{\beta }\right)}{g_\alpha^2 \chi_{\alpha}+g_\beta^2 \chi_{\beta}+\chi_{\mathrm{e}}^{-1}}
\right|^2.
\label{transduction-efficiency-wo-optical-cavity1}
\end{align}
When only one of the two antiferromagnetic magnon modes ($\alpha\ \mathrm{or}\ \beta$) is excited, the transduction efficiency $\eta$ reduces to a simpler expression,
\begin{align}
\eta=\left|
\frac{\sqrt{\kappa _{\mathrm{e,e}}}\sqrt{\xi_\mu} g_\mu}{g_\mu^2+\chi_\mu^{-1}\chi_{\mathrm{e}}^{-1}}
\right|^2
\label{transduction-efficiency-wo-optical-cavity2}
\end{align}
with $\mu=\alpha$ or $\beta$, which takes the same form as in the case of the microwave-to-optical quantum transduction with ferromagnets \cite{Hisatomi2016,Han2021}.

\section{Estimation of the Transduction Efficiency \label{Sec-Numerical-Results}}
So far, we have derived analytical expressions for the transduction efficiency of the microwave-to-optical quantum transduction mediated by antiferromagnetic magnons in antiferromagnets in the cases with and without an optical cavity (where a microwave cavity is used in both cases).
In this section, we estimate the magnitude of the transduction efficiency in the cases with and without an optical cavity.

For concreteness, we consider the case of easy-axis antiferromagnetic insulators such as MnF$_2$, as in the previous sections.
The spin density of MnF$_2$ is $SN/V_{\mathrm{AFM}}\sim10^{19}\, \mathrm{mm}^{-3}$ \cite{Kotthaus1972}.
Here,  $V_{\mathrm{AFM}}$ is the sample volume.
The exchange and anisotropy frequencies of MnF$_2$ are $\omega_E/2\pi=9.3\, \mathrm{THz}$ and $\omega_\parallel/2\pi=0.15\, \mathrm{THz}$, respectively, and the magneto-optic asymmetry between two sublattices of MnF$_2$ is $K=0.007$ \cite{Barak1978,Lockwood2012}.
In the following, we use the above values for MnF$_2$ to obtain the values of $g_\mu$, $\zeta_\mu$, and $\xi_\mu$.

It should be noted that the microwave-magnon interaction strength $g_\mu$ [Eq.~(\ref{microwave-magnon-coupling-strength})] depends explicitly on the frequency of the microwave cavity, i.e., the antiferromagnetic resonance frequency as $\propto\sqrt{\omega_{\mathrm{e}}}$, while the light-magnon interaction strength $\zeta_\mu$ [Eq.~(\ref{light-magnon-coupling-strength-G})] does not.
We take as an example $g_0/2\pi=A\sqrt{\omega_{\mathrm{e}}}$ with $A=25\, \mathrm{mHz}/\sqrt{1\, \mathrm{GHz}}$ \cite{Boventer2023}.
Unless otherwise noted, we also assume that $G_\mu/2\pi=0.1\kappa_\mu/\sqrt{10^9 V_{\mathrm{AFM}}/[\mathrm{mm^3}]}\, \mathrm{MHz}$ \cite{Parvini2020} with $V_{\mathrm{AFM}}$ being given in units of mm$^3$, which is obtained from the values for YIG \cite{Stancil-Book}: $\theta_{\mathrm{F}}=\theta_{\mathrm{F,YIG}}\approx 20^\circ \mathrm{/mm}$, the spin density $n_{\mathrm{YIG}}=2.1\times 10^{19}\, \mathrm{mm^{-3}}$, and $\varepsilon_{\mathrm{r}}\approx 5$.

\subsection{The case with an optical cavity \label{Sec-Estimation-with-optical-cavity}}
Let us consider the case with an optical cavity.
We set $V_{\mathrm{AFM}}=(0.1\, \mathrm{mm})^3$.
First, we assume that one of the two antiferromagnetic magnon modes ($\alpha\ \mathrm{or}\ \beta$) is excited and the triple resonance condition such that $\omega=\delta\omega_{\mathrm{o}}=\omega_\mu=\omega_{\mathrm{e}}$ is satisfied.
Then, Eq.~(\ref{transduction-efficiency-w-optical-cavity2}) gives rise to an expression for the transduction efficiency in terms of the cooperativities,
\begin{align}
\eta=\eta_{\mathrm{o}}\eta_{\mathrm{e}}\frac{4C_{\mathrm{om},\mu}C_{\mathrm{em},\mu}}{(1+C_{\mathrm{om},\mu}+C_{\mathrm{em},\mu})^2},
\label{Transduction-efficiency-w-optical-cavity-resonance}
\end{align}
where $\mu=\alpha$ or $\beta$, $\eta_{\mathrm{o}}=\kappa_{\mathrm{o,e}}/\kappa_{\mathrm{o}}$, $\eta_{\mathrm{e}}=\kappa_{\mathrm{e,e}}/\kappa_{\mathrm{e}}$, and $C_{\mathrm{em},\mu}=4g_\mu^2/(\kappa_{\mathrm{e}}\gamma_\mu)$ and $C_{\mathrm{om},\mu}=4\zeta_\mu^2/(\kappa_{\mathrm{o}}\gamma_\mu)$ are the cooperativity between microwave photons and antiferromagnetic magnons and the cooperativity between optical photons and antiferromagnetic magnons, respectively.
Eq.~(\ref{Transduction-efficiency-w-optical-cavity-resonance}) coincides with a generic expression for the transduction efficiency mediated by one intermediate bosonic mode \cite{Han2021}.
To estimate the magnitude of the transduction efficiency from Eq.~(\ref{Transduction-efficiency-w-optical-cavity-resonance}), suppose that the lower-frequency magnon mode with $\omega_\beta/2\pi=20\, \mathrm{GHz}$ is excited in an easy-axis antiferromagnet\footnote{Note that in the case of easy-axis antiferromagnets a relatively high magnetic field of $\sim 5\, \mathrm{T}$ would be required for realizing a lower-frequency mode of $\sim 20\, \mathrm{GHz}$. On the other hand, in a canted easy-plane antiferromagnet with the Dzyaloshinskii-Moriya interaction, a low magnetic field of $\sim 0.5\, \mathrm{T}$ can be sufficient to realize such a lower-frequency mode \cite{Boventer2023}.}, as experimentally realized \cite{Boventer2023}.
We use the following possible values of parameters: $g_\beta/2\pi\approx 3.3\, \mathrm{MHz}$, $\zeta_\beta/2\pi \approx 40\, \mathrm{kHz}$ (with $\kappa_\beta\approx 0.4$ and $\bar{n}_{\mathrm{cav}}\approx 1\times10^6$), $\gamma_\beta/2\pi\approx 100\, \mathrm{MHz}$, $\kappa_{\mathrm{o,i}}/2\pi=\kappa_{\mathrm{o,e}}/2\pi\approx 100\, \mathrm{MHz}$, and $\kappa_{\mathrm{e,i}}/2\pi=\kappa_{\mathrm{e,e}}/2\pi\approx 100\, \mathrm{MHz}$ \cite{Zhu2020,Parvini2020,Boventer2023}.
Substituting these possible values into Eq.~(\ref{Transduction-efficiency-w-optical-cavity-resonance}), the magnitude of the transduction efficiency is estimated to be
\begin{align}
\eta\sim 10^{-9}, 
\label{Transduction-efficiency-with-lower-magnon}
\end{align}
which is about an order of magnitude smaller than the transduction efficiency utilizing the ferromagnet YIG \cite{Zhu2020}.
This is mainly due to the fact that YIG has a quite small magnon decay rate of $\gamma_{\mathrm{m}}\approx 1\, \mathrm{MHz}$, resulting in a higher transduction efficiency.

\begin{figure}[!t]
\centering
\includegraphics[width=\columnwidth]{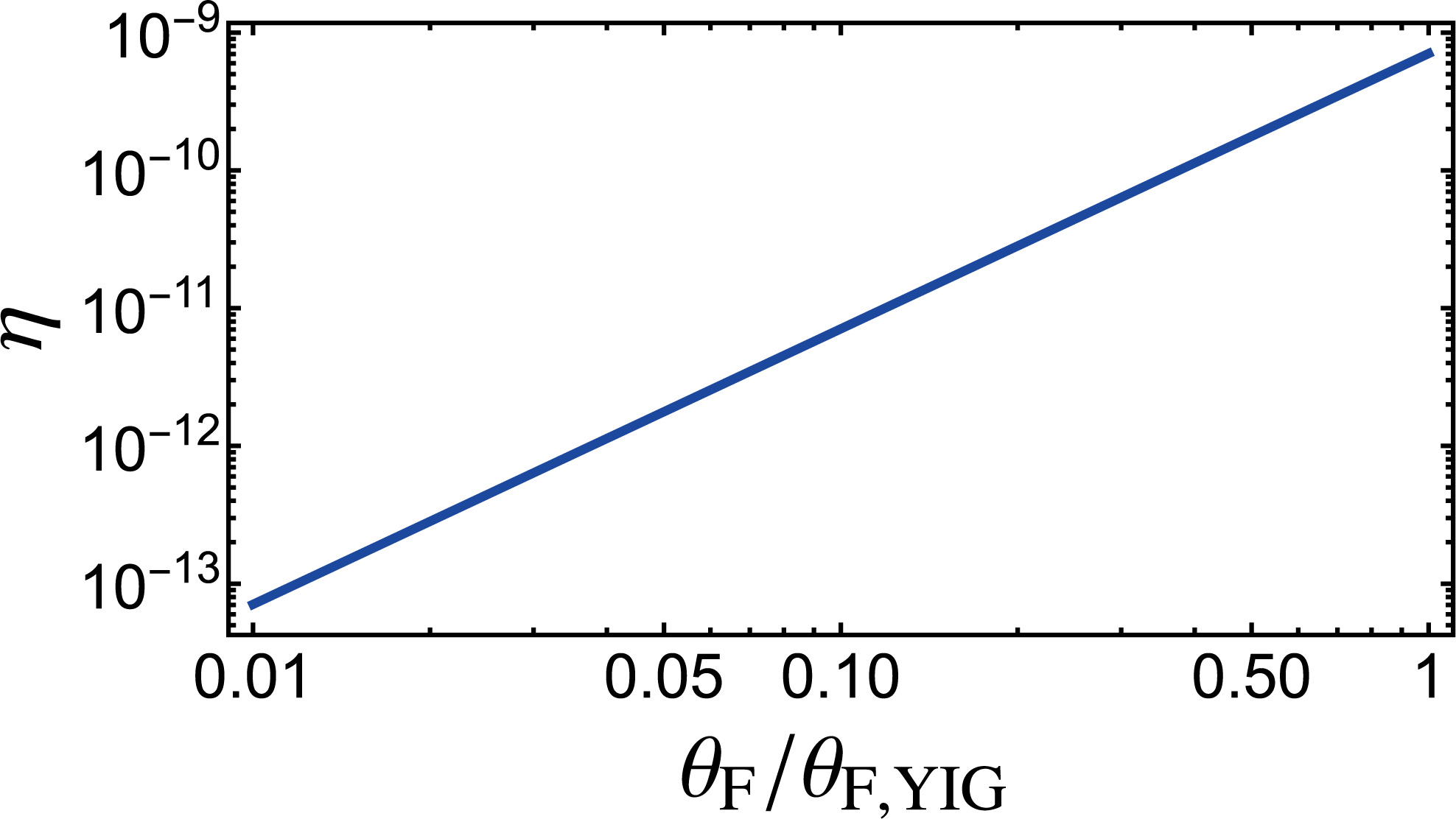}
\caption{Transduction efficiency $\eta$ as a function of the Faraday rotation angle per sublattice (per unit length) $\theta_{\mathrm{F}}$ in the case with an optical cavity.
Here, $\theta_{\mathrm{F,YIG}}$ is the Faraday rotation angle per unit length of the ferromagnet YIG.
Equation~(\ref{Transduction-efficiency-with-lower-magnon}) corresponds to the value with $\theta_{\mathrm{F}}/\theta_{\mathrm{F,YIG}}=1$ in this figure.
}\label{Fig4}
\end{figure}
In the process of estimating the magnitude of the transduction efficiency [Eq.~(\ref{Transduction-efficiency-with-lower-magnon})], we have assumed that $\theta_{\mathrm{F}}/\theta_{\mathrm{F,YIG}}=1$ due to the lack of the experimental data on the value of $\theta_{\mathrm{F}}$ (the Faraday rotation angle per sublattice) in antiferromagnets.
However, this assumption might be inappropriate, since in general the magnitude of the one-magnon Raman scattering is rather small compared to that of the two-magnon Raman scattering in antiferromagnets \cite{Fleury1968,Lockwood1987}.
Therefore, we plot in Fig.~\ref{Fig4} the transduction efficiency $\eta$ as a function of $\theta_{\mathrm{F}}$ (or equivalently, as a function of $G_\beta$).
We find that the transduction efficiency monotonically increases as the value of $\theta_{\mathrm{F}}$ increases.
When $\theta_{\mathrm{F}}/\theta_{\mathrm{F,YIG}}=0.01$, we have $\eta\sim 10^{-13}$.
Because $C_{\mathrm{om},\mu}\ll 1$ in our setup, it follows from Eq.~(\ref{Transduction-efficiency-w-optical-cavity-resonance}) that the tranduction efficiency scales as $\eta\sim\zeta_\beta^2\propto\theta_{\mathrm{F}}^2$.

Next, we turn to the case when the resonance frequencies of the two antiferromagnetic magnon modes are degenerate, which can be realized at zero magnetic field in easy-axis antiferromagnets.
Taking MnF$_2$ as an example, we have $\omega_\alpha/2\pi=\omega_\beta/2\pi\approx 250\, \mathrm{GHz}$ at $B_0=0$ \cite{Rezende2019}.
In this case, it is not easy to obtain a simpler expression even when the quadruple resonance condition $\omega=\delta\omega_{\mathrm{o}}=\omega_\alpha=\omega_\beta=\omega_{\mathrm{e}}$ is satisfied.
Here, it is important to note that both the microwave-magnon interaction strength $g_\mu$ and light-magnon interaction strength $\zeta_\mu$ can be, in general, different depending on the antiferromagnetic magnon mode.
In easy-axis antiferromagnets, which we are focusing on here, we have $\zeta_\alpha\neq\zeta_\beta$ [see Eq.~(\ref{light-magnon-coupling-strength-kappa})] whereas $g_\alpha=g_\beta$ [see Eq.~(\ref{microwave-magnon-coupling-strength})].
For example, it has been estimated that $\kappa_{\alpha,\beta}\approx 0.5,0.4$ [Eq.~(\ref{light-magnon-coupling-strength-kappa})] in the easy-axis antiferromagnet MnF$_{2}$ \cite{Parvini2020}.
We use the following possible values of parameters: $g_\alpha/2\pi=g_\beta/2\pi\approx 10\, \mathrm{MHz}$, $\zeta_\alpha/2\pi \approx 50\, \mathrm{kHz}$, $\zeta_\beta/2\pi \approx 40\, \mathrm{kHz}$ (with $\bar{n}_{\mathrm{cav}}\approx 1\times10^6$), $\gamma_\alpha/2\pi=\gamma_\beta/2\pi\approx 1000\, \mathrm{MHz}$, $\kappa_{\mathrm{o,i}}/2\pi=\kappa_{\mathrm{o,e}}/2\pi\approx 100\, \mathrm{MHz}$, and $\kappa_{\mathrm{e,i}}/2\pi=\kappa_{\mathrm{e,e}}/2\pi\approx 500\, \mathrm{MHz}$ \cite{Zhu2020,Parvini2020,Boventer2023}.
Here, we have assumed that the magnon decay rates ($\gamma_\alpha$ and $\gamma_\beta$) and the microwave cavity decay rates ($\kappa_{\mathrm{e,i}}$ and $\kappa_{\mathrm{e,e}}$) become larger as the antiferromagnetic resonance frequency becomes larger, as in the case of ferromagnets.
Substituting these possible values into Eq.~(\ref{transduction-efficiency-w-optical-cavity1}), the transduction efficiency is estimated to be
\begin{align}
\eta\sim 10^{-10}.
\label{Transduction-efficiency-with-two-magnon}
\end{align}
Note that the smallness of this value compared to Eq.~(\ref{Transduction-efficiency-with-lower-magnon}) is due to the above assumption that the decay rates become larger as the antiferromagnetic resonance frequency become larger.
Similar magnitudes of the transduction efficiency to Eq.~(\ref{Transduction-efficiency-with-two-magnon}) are also obtained when only the higher-frequency magnon mode with $\omega_\alpha/2\pi\sim 10^2\, \mathrm{GHz}$ is excited.

\subsection{The case without an optical cavity \label{Sec-Estimation-without-optical-cavity}}
Let us consider the case without an optical cavity.
Recall that the thickness of the antiferromagnet  $d_{\mathrm{AFM}}$ should be of, or shorter than, $\mathcal{O}(1\, \mathrm{\mu m})$ (see Sec.~\ref{Sec-Light-magnon-wo-optical-cavity}).
Hence, we set $V_{\mathrm{AFM}}=(0.1\, \mathrm{mm})^2\times 1\, \mathrm{\mu m}$ with $d_{\mathrm{AFM}}=1\, \mathrm{\mu m}$.
First, we assume that one of the two antiferromagnetic magnon modes ($\alpha\ \mathrm{or}\ \beta$) is excited and the double resonance condition such that $\omega=\omega_\mu=\omega_{\mathrm{e}}$ is satisfied.
Then, Eq.~(\ref{transduction-efficiency-wo-optical-cavity2}) gives rise to an expression for the transduction efficiency in terms of the cooperativity,
\begin{align}
\eta=\eta_{\mathrm{e}}\eta_{\mathrm{m},\mu}\frac{4C_{\mathrm{em},\mu}}{(1+C_{\mathrm{em},\mu})^2},
\label{Transduction-efficiency-wo-optical-cavity-resonance}
\end{align}
where $\mu=\alpha$ or $\beta$, $\eta_{\mathrm{m},\mu}=\xi_\mu/\gamma_\mu$, and the other quantities have been defined in Eq.~(\ref{Transduction-efficiency-w-optical-cavity-resonance}).
As in the case with an optical cavity, Eq.~(\ref{Transduction-efficiency-wo-optical-cavity-resonance}) coincides with a generic expression for the transduction efficiency mediated by one intermediate bosonic mode \cite{Han2021}.
To estimate the magnitude of the transduction efficiency from Eq.~(\ref{Transduction-efficiency-wo-optical-cavity-resonance}), suppose that the lower-frequency magnon mode with $\omega_\beta/2\pi=20\, \mathrm{GHz}$ is excited in an easy-axis antiferromagnet, as experimentally realized \cite{Boventer2023}.
We use the following possible values of parameters: $g_\beta/2\pi\approx 0.33\, \mathrm{MHz}$, $\gamma_\beta/2\pi\approx 100\, \mathrm{MHz}$, and $\kappa_{\mathrm{e,i}}/2\pi=\kappa_{\mathrm{e,e}}/2\pi\approx 100\, \mathrm{MHz}$ \cite{Boventer2023}.
As for the light-magnon interaction strength [Eq.~(\ref{Light-magnon-interaction-without-cavity-strength})], we assume that $P_0=15\, \mathrm{mW}$ and $\Omega_0/2\pi=193\, \mathrm{THz}$, which gives rise to $\xi_\beta/2\pi\approx 2.1\times 10^{-7}\, \mathrm{Hz}$.
Substituting these possible values into Eq.~(\ref{Transduction-efficiency-wo-optical-cavity-resonance}), the transduction efficiency is estimated to be
\begin{align}
\eta\sim 10^{-19}.
\label{Transduction-efficiency-with-lower-magnon-wo-optical-cavity}
\end{align}
The smallness of this value results from the small light-magnon interaction strength $\xi_\beta$, i.e., the smallness of $\eta_{\mathrm{m},\beta}\sim 10^{-15}$, as can be seen that the transduction efficiency $\eta$ is proportional to $\eta_{\mathrm{m},\beta}$ in Eq.~(\ref{Transduction-efficiency-wo-optical-cavity-resonance}).

When the resonance frequencies of the two antiferromagnetic magnon modes are degenerate, or when the higher-frequency magnon mode with $\omega_\alpha/2\pi\sim 10^2\, \mathrm{GHz}$ is excited, the magnon decay rates $\gamma_\alpha$ and $\gamma_\beta$ would be larger than those in the frequency range of $\omega_\mu\sim 10^1\, \mathrm{GHz}$ (see Sec.~\ref{Sec-Estimation-with-optical-cavity}).
Then, the magnitude of the transduction efficiency would become lower than Eq.~(\ref{Transduction-efficiency-with-lower-magnon-wo-optical-cavity}).
This is because the magnitude of the transduction efficiency in the form of Eq.~(\ref{Transduction-efficiency-with-lower-magnon-wo-optical-cavity}) is essentially determined by the magnitude of $\eta_{\mathrm{m},\beta}$, as has been discussed in the case of ferromagnets \cite{Hisatomi2016,Sekine2024}.

\subsection{Thickness dependence}
Here, we investigate the thickness dependence of the transduction efficiency in the cases with and without an optical cavity.
To this end, we parametrize the sample thickness $d_{\mathrm{AFM}}\, [\mathrm{mm}]$ and set the sample volume as $V_{\mathrm{AFM}}=(0.1\, \mathrm{mm})^2\times d_{\mathrm{AFM}}$.
Then, assuming that the lower-frequency magnon mode with $\omega_\beta/2\pi=20\, \mathrm{GHz}$ is excited in an easy-axis antiferromagnet and using the parameters introduced in Secs.~\ref{Sec-Estimation-with-optical-cavity} and \ref{Sec-Estimation-without-optical-cavity}, we obtain
\begin{align}
g_\beta&\approx 10.5\times\sqrt{\frac{d_{\mathrm{AFM}}}{[\mathrm{mm}]}}\, \mathrm{MHz},\nonumber\\
\zeta_\beta&\approx 1.3\times 10^{-2}/\sqrt{\frac{d_{\mathrm{AFM}}}{[\mathrm{mm}]}}\, \mathrm{MHz},\nonumber\\
\xi_\beta&\approx 2.1\times10^{-10}\times \frac{d_{\mathrm{AFM}}}{[\mathrm{mm}]}\, \mathrm{MHz}.
\label{interaction-strengths-thickness-dependence}
\end{align}
Here, note that one can consider directly the sample volume $V_{\mathrm{AFM}}$ dependence instead of the thickness dependence.
Of course, these two parameters are essentially equivalent.

\begin{figure}[!t]
\centering
\includegraphics[width=\columnwidth]{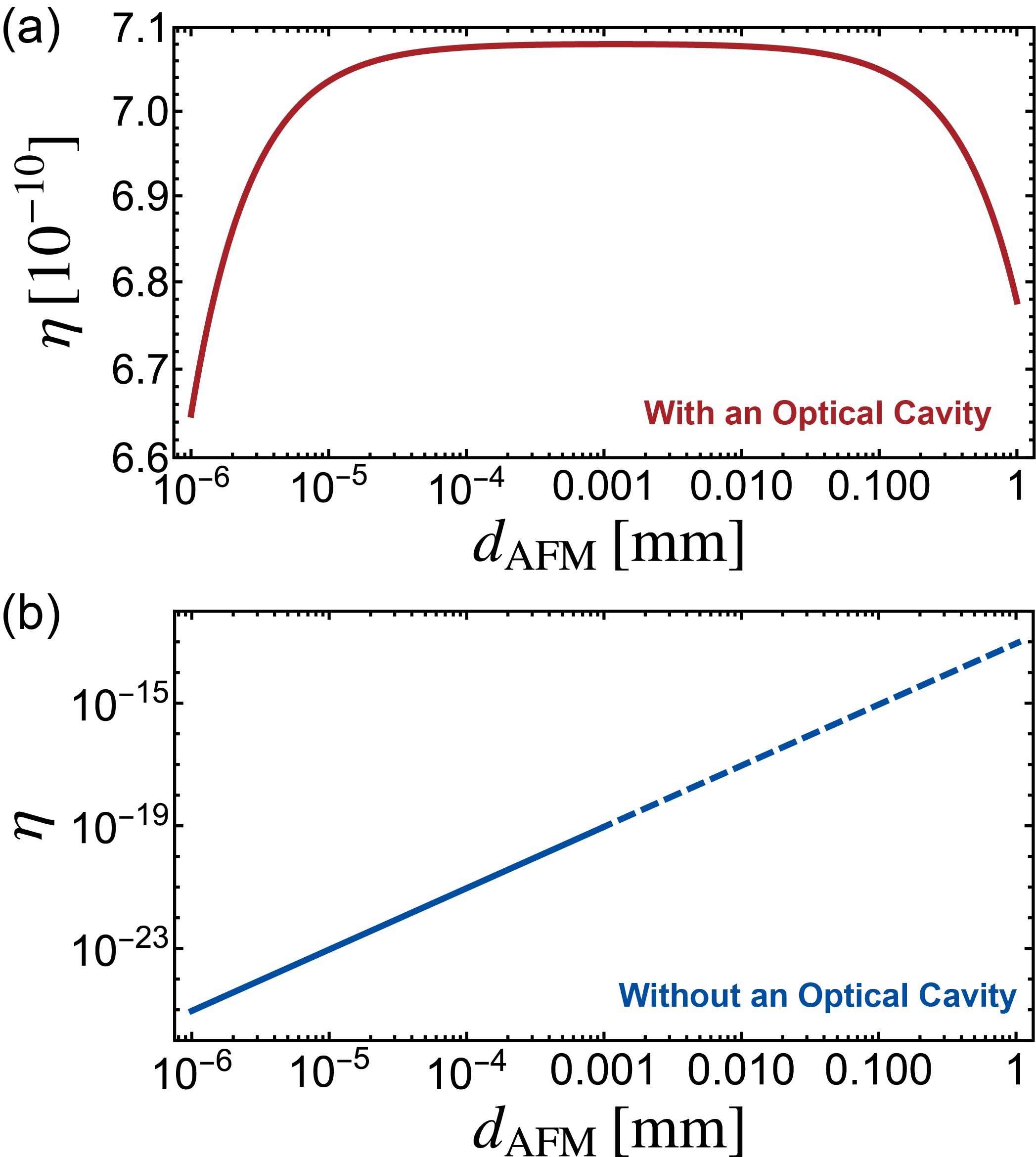}
\caption{(a) Thickness $d_{\mathrm{AFM}}$ dependence of the transduction efficiency $\eta$ in the case with an optical cavity.
(b) Thickness $d_{\mathrm{AFM}}$ dependence of the transduction efficiency $\eta$ in the case without an optical cavity.
Note that in (b) the transduction efficiency $\eta$ is defined when the thickness $d_{\mathrm{AFM}}$ is of, or shorter than, $\mathcal{O}(10^{-3}\, \mathrm{mm})$, whereas in (a) there is no restriction on the value of $d_{\mathrm{AFM}}$.
}\label{Fig5}
\end{figure}
As shown in Fig.~\ref{Fig5}, we find a distinct behavior of the transduction efficiency $\eta$ as a function of the thickness $d_{\mathrm{AFM}}$ depending on the presence or absence of an optical cavity.
First, we show in Fig.~\ref{Fig5}(a) the result for the case with an optical cavity, which is obtained from Eq.~(\ref{Transduction-efficiency-w-optical-cavity-resonance}).
In this case, we see that there exists an optimal value of the thickness $d_{\mathrm{AFM}}\sim \mathcal{O}(10^{-3}\, \mathrm{mm})$ at which the transduction efficiency takes a maximum value.
This result can be understood as follows.
When the value of $d_{\mathrm{AFM}}$ is small, the effect of the light-magnon interaction $\zeta_\beta\propto 1/\sqrt{d_{\mathrm{AFM}}}$, which is an decreasing function of $d_{\mathrm{AFM}}$, is dominant, and thus the transduction efficiency scales as $\eta\sim C_{\mathrm{om},\beta}^{-1}\sim d_{\mathrm{AFM}}$.
On the other hand, when the value of $d_{\mathrm{AFM}}$ is large, the effect of the microwave-magnon interaction $\xi_\beta\propto \sqrt{d_{\mathrm{AFM}}}$, which is an increasing function of $d_{\mathrm{AFM}}$, is dominant, and thus the transduction efficiency scales as $\eta\sim C_{\mathrm{em},\beta}^{-1}\sim d_{\mathrm{AFM}}^{-1}$.

Next, we show in Fig.~\ref{Fig5}(b) the result for the case without an optical cavity, which is obtained from Eq.~(\ref{Transduction-efficiency-wo-optical-cavity-resonance}).
In this case, we see that the transduction efficiency is a monotonically increasing function of the thickness $d_{\mathrm{AFM}}$.
This result can be understood from that both the light-magnon interaction $\xi_\beta$ and the microwave-magnon interaction $g_\beta$ are increasing functions of $d_{\mathrm{AFM}}$, and thus the transduction is also an increasing function of $d_{\mathrm{AFM}}$.

Finally, we note that the above distinct behavior of the transduction efficiency with respect to the sample thickness depending on the presence or absence of an optical cavity will also be observed when utilizing ferromagnets, since both the microwave-magnon and light-magnon interaction strengths in ferromagnets have the same sample thickness dependences as Eq.~(\ref{interaction-strengths-thickness-dependence}) \cite{Hisatomi2016,Kusminskiy2016,Sekine2024} [see also Eq.~(\ref{microwave-magnon-coupling-strength}) and Eqs.~(\ref{light-magnon-coupling-strength-G}) and (\ref{Light-magnon-interaction-without-cavity-strength}), respectively].

\section{Discussion \label{Sec-Discussion}}
\begin{figure}[!t]
\centering
\includegraphics[width=\columnwidth]{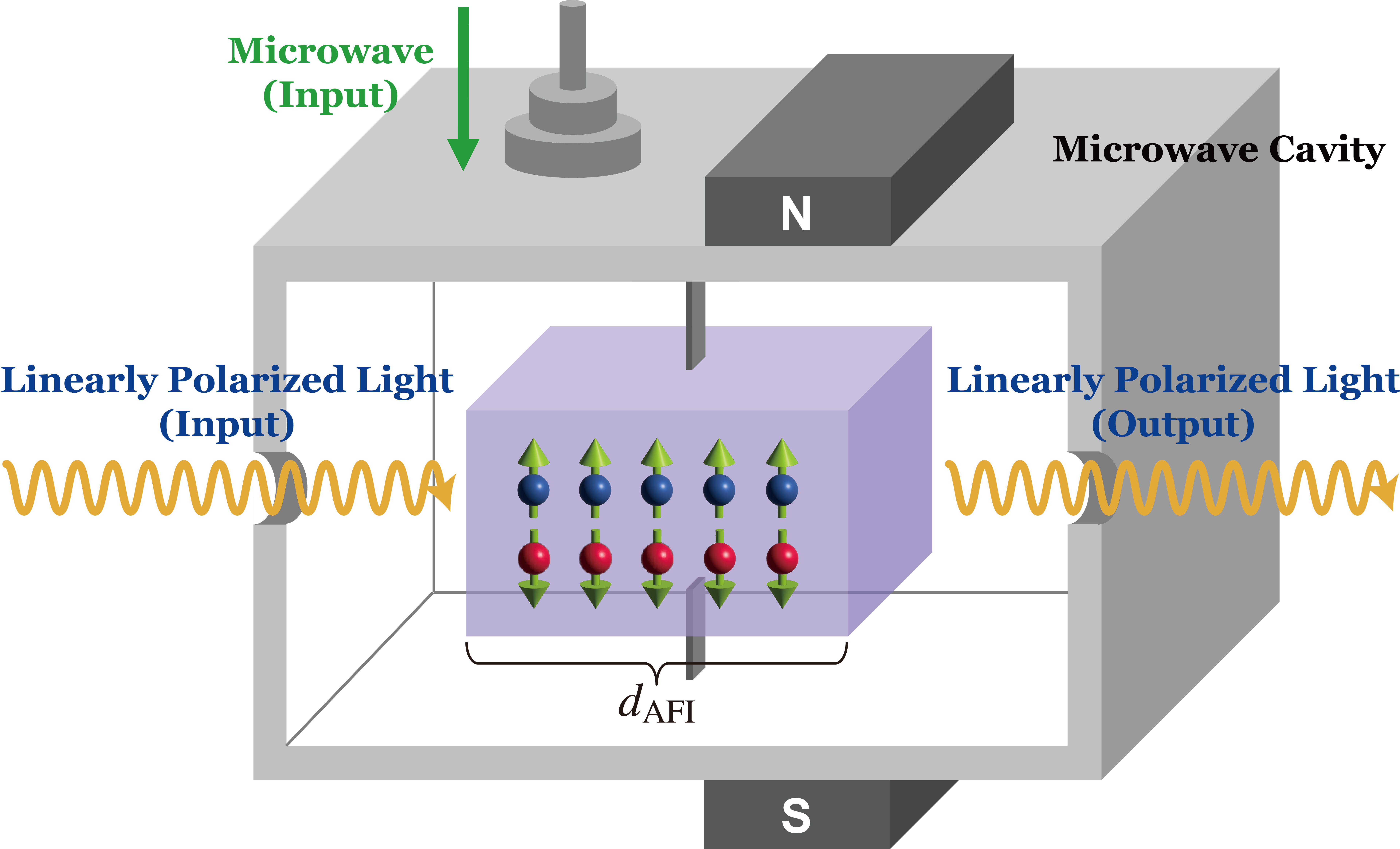}
\caption{Schematic illustration of our experimental setup for the microwave-to-optical quantum transduction without an optical cavity.
A static magnetic field is applied along the easy axis of the antiferromagnet.
A linearly polarized laser with the frequency of $193\, \mathrm{THz}$ is applied perpendicular to the easy axis of the antiferromagnet.
}\label{Fig6}
\end{figure}

Let us briefly discuss a possible experimental setup of the microwave-to-optical quantum transduction utilizing antiferromagnets.
A schematic illustration of our experimental setup for the case without an optical cavity is shown in Fig.~\ref{Fig6}.
For concreteness, we show the case of an easy-axis antiferromagnet where only one of the two antiferromagnetic magnon modes ($\alpha$ or $\beta$) is excited.
A static magnetic field is applied along the easy axis and a linearly polarized laser with the frequency of $193\, \mathrm{THz}$ is applied perpendicular to the easy axis.
On the other hand, when we take into account an optical cavity, we can for example use an optical waveguide as has been used for the microwave-to-optical quantum transduction utilizing the ferromagnet YIG \cite{Zhu2020}.
As we have considered in Sec.~\ref{Sec-Light-magnon-with-optical-cavity}, the light needs be applied perpendicular to the easy axis.

One of the merits of utilizing antiferromagnets for the microwave-to-optical quantum transduction is the wide frequency tunability of the microwave cavity frequency (corresponding to the antiferromagnetic resonance frequency), ranging from $\mathcal{O}(1\, \mathrm{GHz})$ to $\mathcal{O}(1\, \mathrm{THz})$.
This means that a variety of quantum devices that operate at these frequencies can be interconnected via the quantum transducer utilizing antiferromagnets.
Another merit would be the operating temperature of $\mathcal{O}(100\, \mathrm{K})$ of the quantum transduction higher than those of other transduction methods using the optomechanical and electro-optic effects \cite{Lauk2020,Lambert2020,Han2021}, originating from the fact the antiferromagnetic resonance can occur at $\mathcal{O}(100\, \mathrm{K})$.
(Note that this merit also applies to the case of ferromagnets.)

\begin{figure}[!t]
\centering
\includegraphics[width=0.7\columnwidth]{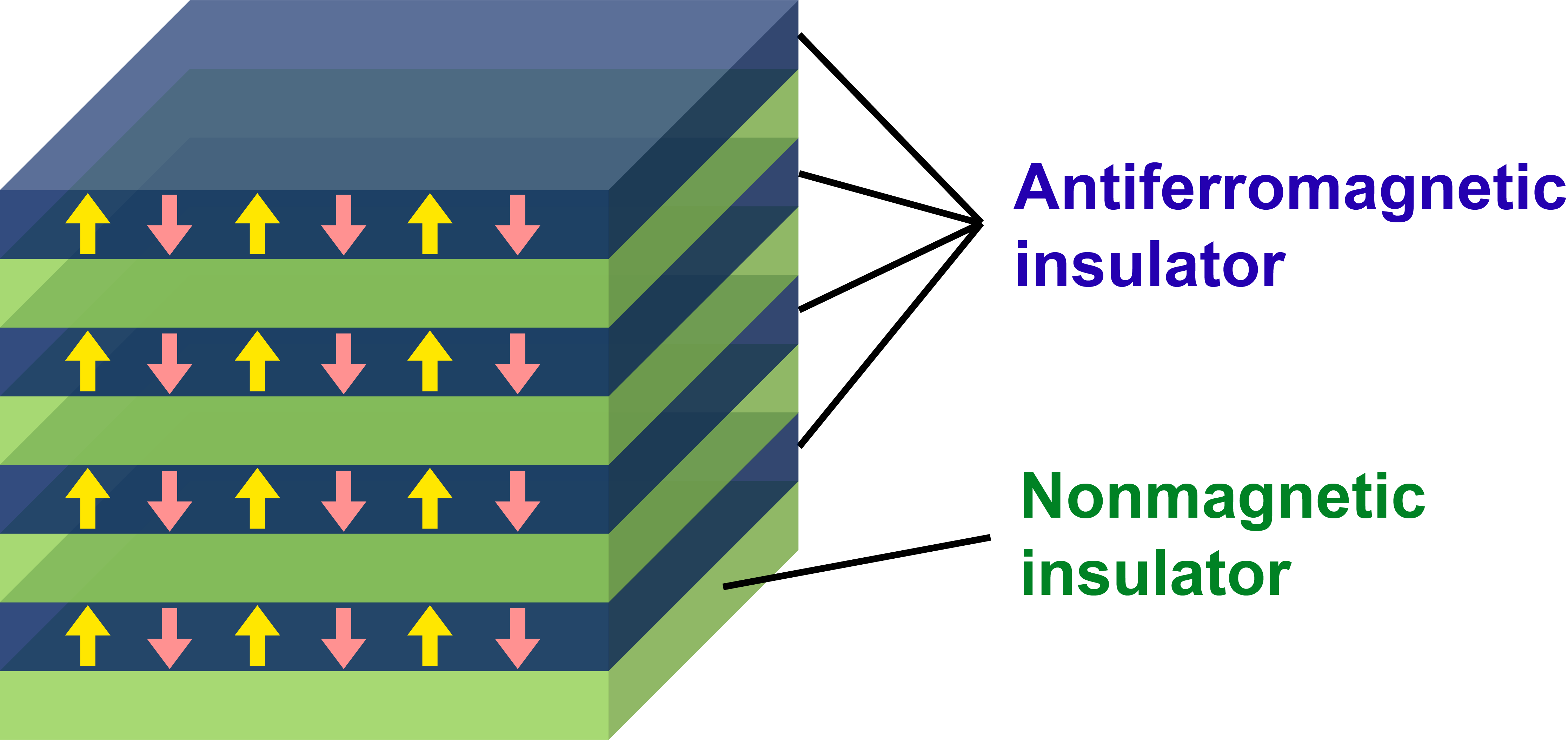}
\caption{Schematic illustration of a heterostructure consisting of $N_{\mathrm{L}}$ antiferromagnetic insulator layers and $N_{\mathrm{L}}$ nonmagnetic insulator layers.
}\label{Fig7}
\end{figure}

The magnitudes of the transduction efficiency we have estimated so far are not sufficient for practical use as a quantum transducer, since it has been suggested theoretically that $\eta=0.5$ is the lower limit for realizing the quantum state transfer between quantum devices \cite{Lauk2020,Lambert2020,Han2021}.
Let us consider possible ways of improving the transduction efficiency.
As we have seen in Secs.~\ref{Sec-Estimation-with-optical-cavity} and \ref{Sec-Estimation-without-optical-cavity}, in both cases with and without an optical cavity the transduction efficiency gets improved as the magnon decay rate becomes smaller.
Thus, utilizing antiferromagnets with a low magnon decay rate is one possible direction.
Another direction may be to utilize the heterostructures consisting of an antiferromagnetic insulator and a nonmagnetic insulator, in a similar way to the topological insulator heterostructures \cite{Sekine2024}.
Here, let us briefly take a look at the case with an optical cavity, because there is a sample size limitation in the case without an optical cavity whereas there is no limitation in the case with an optical cavity (see Sec.~\ref{Sec-Light-magnon-wo-optical-cavity}).
For a heterostructure of $N_{\mathrm{L}}$ antiferromagnetic insulator layers and $N_{\mathrm{L}}$ nonmagnetic insulator layers schematically shown in Fig.~\ref{Fig7}, the microwave-magnon and light-magnon interaction strengths are modified as $g_\mu\to\tilde{g}_\mu=g_\mu\sqrt{N_{\mathrm{L}}}$ and $\zeta_\mu\to\tilde{\zeta}_\mu=\zeta_\mu\sqrt{N_{\mathrm{L}}}$, respectively\footnote{Note that the total interaction Hamiltonians of a heterostructure will be $H_g^{\mathrm{tot}}=\sum_{i=1}^{N_{\mathrm{L}}}\sum_{\mu=\alpha,\beta}\hbar g_{i,\mu}\left(\hat{a}^\dag \hat{m}_{i,\mu}+\hat{m}^\dag_{i,\mu} \hat{a}\right)$ and $H_\zeta^{\mathrm{tot}}=-\sum_{i=1}^{N_{\mathrm{L}}}\sum_{\mu=\alpha,\beta}\hbar\zeta_{i,\mu}\left(\hat{m}_{i,\mu}^\dag+\hat{m}_{i,\mu}\right)\left(\hat{b}^\dag+\hat{b}\right)$. Assuming an equal volume (or thickness) of each antiferromagnetic insulator layer, we can set $ g_{i,\mu}\equiv g_\mu$ and $\zeta_{i,\mu}\equiv \zeta_\mu$.}.
This is because the magnon operators in Eqs.~(\ref{Hamiltonian-H_g}) and (\ref{Light-magnon-interaction-with-cavity}) are modified as $\hat{m}_\mu\to\hat{\mathfrak{m}}_\mu=\frac{1}{\sqrt{N_{\mathrm{L}}}}\sum_{i=1}^{N_{\mathrm{L}}} \hat{m}_{i,\mu}$ ($\mu=\alpha,\beta$), where $\hat{m}_{i,\mu}$ is the $\mu$-mode magnon operator of the $i$-th antiferromagnetic insulator layer and $\hat{\mathfrak{m}}_\mu$ is the {\it collective} magnon operator satisfying the commutation relation $[\hat{\mathfrak{m}}_\mu,\hat{\mathfrak{m}}_\mu^\dag]=1$ \cite{Sekine2024}.
Then, we find that the transduction efficiency $\eta$ [Eq.~(\ref{Transduction-efficiency-w-optical-cavity-resonance})] in the triple resonance condition scales as $\eta\sim N_{\mathrm{L}}^2$, since with the parameters we have used in this study the cooperativities are small so that $\eta\sim C_{\mathrm{om},\mu}C_{\mathrm{em},\mu}$ in Eq.~(\ref{Transduction-efficiency-w-optical-cavity-resonance}).
If a heterostructure with $N_{\mathrm{L}}=5000$ and $d_{\mathrm{AFM}}=1\, \mathrm{\mu m}$ (the thickness of each layer) is possible, then the transduction efficiency is improved from $\eta\sim 10^{-9}$ [Eq.~(\ref{Transduction-efficiency-with-lower-magnon})] to
\begin{align}
\eta\sim 10^{-2},
\end{align}
which is comparable to the current achievements in optomechanical and electro-optic systems \cite{Lauk2020,Lambert2020,Han2021}.

Finally, we mention a recent study on a microwave-to-optical quantum transduction using antiferromagnetic topological insulators \cite{Xu2024}.
The quantum transduction studied in Ref.~\cite{Xu2024} is based on a strong nonlinear optical interaction via spin-orbit coupling, which gives rise to the effective Hamiltonian of the form $H_{\mathrm{int}}=\hbar\mathcal{G}(\hat{a}^\dag\hat{b}+\hat{b}^\dag\hat{a})$, where $\hat{a}$ and $\hat{b}$ are the annihilation operator of microwave and optical photons, respectively.
This mechanism is similar to that of the quantum transduction via the electro-optic effect, which is thus different from ours, i.e., the quantum transduction via the intermediate bosonic mode.

\section{Summary \label{Sec-Summary}}
To summarize, we have formulated a theory for the microwave-to-optical quantum transduction mediated by antiferromagnetic magnons in antiferromagnets.
We have derived analytical expressions for the transduction efficiency in the cases with and without an optical cavity, where a microwave cavity is used in both cases.
The derived expressions for the transduction efficiency [Eqs.~(\ref{transduction-efficiency-w-optical-cavity1}) and (\ref{transduction-efficiency-wo-optical-cavity1})] take complicated forms when the two antiferromagnetic magnon modes coexist, while they reduce to the same expressions as in the case utilizing ferromagnets when one of the two antiferromagnetic magnon modes are excited. 
Here, we stress that the derived expressions for the transduction efficiency [Eqs.~(\ref{transduction-efficiency-w-optical-cavity1}) and (\ref{transduction-efficiency-wo-optical-cavity1})] cannot be written simply as the sum of the transduction efficiency of each antiferromagnetic mode, which implies that the extension of the ferromagnetic case to the antiferromagnetic case is nontrivial.
We have found that the quantum transduction can occur even in the absence of an external static magnetic field, which is in contrast to the case of the quantum transduction using ferromagnets where an external static magnetic field is essential to cause the resonance state.
We have also found that, in the case with an optical cavity the transduction efficiency takes a peak structure with respect to the sample thickness, indicating that there exists an optimal thickness, whereas in the case without an optical cavity the transduction efficiency is a monotonically increasing function of the sample thickness.
It is expected that by utilizing a heterostructure consisting of an antiferromagnet and a nonmagnetic insulator the transduction efficiency can be improved to the magnitude that is comparable to the up-to-date values achieved so far in optomechanical and electro-optic systems.
Owing to the wide tunability of the antiferromagnetic resonance frequency, it is expected that a variety of quantum devices that operate at the wide frequency ranges from $\mathcal{O}(1\, \mathrm{GHz})$ to $\mathcal{O}(1\, \mathrm{THz})$ can be interconnected via the quantum transducer utilizing antiferromagnets.
Our study opens up a way for possible applications of antiferromagnetic materials in future quantum interconnects.

\acknowledgements
We would like to thank Shintaro Sato and Mari Ohfuchi for their advice and support.

\appendix
\section{Derivation of the light-magnon interaction in antiferromagnets \label{Appendix-Derivation}}
In this Appendix, we outline the derivation of the light-magnon interaction in antiferromagnets [Eqs.~(\ref{Hamiltonian-Parvini2020}) and (\ref{light-magnon-coupling-strength-G})], following Refs.~\cite{Parvini2020,Cottam1975,Cottam-Lockwood-Book}.
The Hamiltonian describing the interaction between light and magnon comes from the electromagnetic energy $U=\int d^3r\, \bm{E}\cdot\bm{D}/2$, where $\bm{D}=\hat{\varepsilon}\bm{E}$ with $\hat{\varepsilon}_{ij}$ the permittivity tensor and $\bm{E}$ an external electric field.
Since the permittivity tensor can be spatial dependent, we have in general \cite{LL-Book}
\begin{align}
H_{\mathrm{int}}&=\frac{1}{4}\int d^3r\, \sum_{\mu,\nu}E^*_\mu(\bm{r})\varepsilon_{\mu\nu}(\bm{r})E_\mu(\bm{r})\nonumber\\
&=\frac{V}{4N}\sum_i\sum_{\mu,\nu}E^*_\mu(\bm{r}_i)\varepsilon_{\mu\nu}(\bm{r}_i)E_\mu(\bm{r}_i),
\label{Generic-interaction-Hamiltonian}
\end{align}
where $\mu,\nu=x,y,z$ denote a spatial direction, $V$ is the system volume, $N$ is the total number of lattice sites, and $i$ runs over the all lattice cites.
In magnetic systems, the permittivity tensor can be expanded in terms of spins on lattice cites.
Up to the terms quadratic in spin operators, we get \cite{Moriya1967,Cottam1975,Cottam-Lockwood-Book}
\begin{align}
\varepsilon_{\mu\nu}(\bm{r}_i)=\sum_\rho K_{\mu\nu\rho}S^\rho_i+\sum_{\alpha,\beta} G_{\mu\nu\alpha\beta}S^\alpha_i S^\beta_i+\sum_{\alpha,\beta} B_{\mu\nu\alpha\beta}S^\alpha_i S^\beta_{i+\delta}.
\label{permittivity-expansion}
\end{align}
For antiferromagnets with a rutile crystal structure such as FeF$_2$ and MnF$_2$ which we are interested in, the third term in the right-hand side is very small such that $B_{\mu\nu\alpha\beta}/G_{\mu\nu\alpha\beta}\sim J/\lambda\ll 1$ \cite{Moriya1967}, where $J$ is the exchange coupling strength and $\lambda$ is the spin-orbit coupling strength.

In the present study, we are focusing on the one-magnon scattering process in antiferromagnets.
For this purpose, the second term ($\propto S^z S^\pm$) in the right-hand side of Eq.~(\ref{permittivity-expansion}) can be absorbed into the first term ($\propto S^\pm$).
Thus, in the following, we shall take into account only the first term in Eq.~(\ref{permittivity-expansion}) when deriving the light-magnon interaction.

\subsection{The case of ferromagnets}
For comparison, let us firstly consider the light-magnon interaction in ferromagnets.
In the case of ferromagnets, which can be characterized by their macroscopic magnetization $\bm{M}$, the permittivity tensor is written as a function of the magnetization.
To linear order in the magnetization, we have [see Eq.~(\ref{permittivity-expansion})] \cite{LL-Book}
\begin{align}
\varepsilon_{\mu\nu}(\bm{M})=\varepsilon_0\left(\varepsilon_{\mathrm{r}}\delta_{\mu\nu}-iK_0\sum_\rho\epsilon_{\mu\nu\rho}M_\rho\right),
\label{permittivity-ferromagnets}
\end{align}
where $\varepsilon_0$ ($\varepsilon_{\mathrm{r}}$) is the vacuum (relative) permittivity, $\epsilon_{\mu\nu\rho}$ is the Levi-Civita symbol, and $K_0$ is a material-dependent constant.
Then, from Eqs.~(\ref{Generic-interaction-Hamiltonian}) and (\ref{permittivity-ferromagnets}), we obtain
\begin{align}
H_{\mathrm{int}}=-i\frac{\varepsilon_0 K_0}{4}\int d^3r\, \bm{M}(\bm{r})\cdot\left[\bm{E}^*(\bm{r})\times\bm{E}(\bm{r})\right].
\end{align}
The electric field in an optical cavity can be quantized as $\bm{E}(\bm{r},t)=\sum_\alpha\bm{E}_\alpha(\bm{r})\hat{b}_\alpha(t)$, where $\bm{E}_\alpha(\bm{r})$ is the $\alpha$-th eigenmode of the electric field and $\hat{b}_\alpha(t)$ is the annihilation operator of the photon in the $\alpha$-th eigenmode.

For concreteness, suppose that the magnetization in the ground state is along the $z$ axis and the magnetic moments are precessing around the $z$ axis in the ferromagnetic resonance state.
In this case, the electric field propagating in the direction perpendicular to the $z$ axis interact with the fluctuating component of the magnetic moments, i.e., the ferromagnetic magnon.
We also have the relation $M_i/M_{\mathrm{s}}=\hat{S}_i/(SN)$ with $M_{\mathrm{s}}$ being the saturation magnetization and $SN$ being the total spin number.
Finally, we obtain the light-magnon interaction in ferromagnets \cite{Kusminskiy2016}
\begin{align}
H_{\mathrm{int}}=\hbar G_{\mathrm{FM}} \left(\hat{m}+\hat{m}^\dag\right)\left(\hat{b}_R^\dag\hat{b}_R-\hat{b}_L^\dag\hat{b}_L\right),
\end{align}
where $\hat{m}$ is the annihilation operator for the ferromagnetic magnon mode.
The coupling strength is given by
\begin{align}
G_{\mathrm{FM}}=\frac{c\theta_{\mathrm{F}}}{4\sqrt{\varepsilon_{\mathrm{r}}}}\frac{1}{\sqrt{2SN}},
\end{align}
with the Faraday rotation angle per unit length $\theta_{\mathrm{F}}=\omega_{\mathrm{c}}K_0 M_{\mathrm{s}}/(2c\sqrt{\varepsilon_{\mathrm{r}}})$ ($\omega_{\mathrm{c}}$ is the cavity mode frequency).

\subsection{The case of antiferromagnets}
Now, let us consider the case of antiferromagnets.
For concreteness, we focus on the antiferromagnets with a rutile crystal structure such as FeF$_2$, MnF$_2$, and CoF$_2$ \cite{Cottam1975,Lockwood2012}.
In these antiferromagnets, the antiferromagnetic spins are aligned along the $z$ (crystal $c$) axis in the ground states.
The magnetic ions form a body-centered tetragonal structure, with each magnetic ion surrounded by six non-magnetic ions that form a distorted octahedron.
The two sublattices $A$ and $B$ (the one occupying the body-centered sites and the one occupying the corner sites) are distinguished by a 90$^\circ$ rotation about the $z$ axis.
We choose $x$, $y$, and $z$ axes so that the magnetic ions have a two-fold rotational symmetry with respect to each axis.
Then, from a point-group symmetry analysis, it turns out that for sublattice $A$ \cite{Cottam1975}
\begin{align}
&K_{yzx}^A=-K_{zyx}^A\equiv K_1,\nonumber\\
&K_{zxy}^A=-K_{xzy}^A\equiv K_2,\nonumber\\
&K_{xyz}^A=-K_{yxz}^A\equiv K_3.
\end{align}
Since sublattice $B$ is related to sublattice $A$ by a 90$^\circ$ rotation about the $z$ axis, we have the relation $K_{\mu\nu\rho}^B=\sum_{\mu',\nu',\rho'}R_{\mu\mu'}R_{\nu\nu'}R_{\rho\rho'}K_{\mu'\nu'\rho'}^A$, where $R_{\alpha\beta}$ is a matrix element of the 90$^\circ$ rotation matrix about the $z$ axis.
Therefore, for sublattice $B$,
\begin{align}
&K_{yzx}^B=-K_{zyx}^B=-K_{xzy}^A=K_2,\nonumber\\
&K_{zxy}^B=-K_{xzy}^B=-K_{zyx}^A=K_1,\nonumber\\
&K_{xyz}^B=-K_{yxz}^B=-K_{yxz}^A=K_3.
\end{align}
Note that $K_1$, $K_2$, and $K_3$ are purely imaginary.

As we have discussed briefly, we are focusing on the one-magnon scattering process.
This means that the spin operators $S^x_i$ and $S^y_i$ are relevant, while $S^z_i$ is not because $S^z_i$ is quadratic in the magnon operators $\hat{a}$ and $\hat{b}$. 
Defining $K_\pm=i(K_1\pm K_2)/4$, Eq.~(\ref{Generic-interaction-Hamiltonian}) reduces to \cite{Parvini2020,Cottam1975,Cottam-Lockwood-Book}
\begin{align}
H_{\mathrm{int}}=&\ \frac{K_+ V}{4N}\sum_{i\in A,B}\left(P_i^+S_i^- - P_i^-S_i^+\right)\nonumber\\
&+\frac{K_- V}{4N}\left[\sum_{i\in A}\left(P_i^+S_i^+ - P_i^-S_i^-\right)-\sum_{j\in B}\left(P_j^+S_j^+ - P_j^-S_j^-\right)\right],
\label{Hamiltonian-Light-Spin}
\end{align}
where $P_i^\pm=E^*_z(\bm{r}_i)E_\pm(\bm{r}_i)-E^*_\mp(\bm{r}_i)E_z(\bm{r}_i)$ with $E_\pm=E_x\pm iE_y$.
Finally, by substituting an expression for the quantized electric field in an optical cavity into Eq.~(\ref{Hamiltonian-Light-Spin}), we arrive at the expression for the light-magnon interaction, which  corresponds to the one-magnon scattering, in the main text [Eqs.~(\ref{Hamiltonian-Parvini2020}) and (\ref{Light-magnon-interaction-without-cavity-1})]:
\begin{align}
H_\zeta=-\hbar \left(\hat{b}_R^\dag\hat{b}_R-\hat{b}_L^\dag\hat{b}_L\right)\left[G_\alpha\left(\hat{m}_\alpha^\dag+\hat{m}_\alpha\right)+G_\beta\left(\hat{m}_\beta^\dag+\hat{m}_\beta\right)\right].
\end{align}
Here, the coupling strength is given by
\begin{align}
G_\mu=\frac{c\theta_{\mathrm{F}}}{4\sqrt{\varepsilon_{\mathrm{r}}}}\frac{\kappa_\mu}{\sqrt{2SN}},
\end{align}
with the Faraday rotation angle per unit length $\theta_{\mathrm{F}}=\omega_{\mathrm{c}}K_+ S/(c\sqrt{\varepsilon_{\mathrm{r}}})$.
Since $K_+$ (i.e., $K_1$ and $K_2$) is defined as a quantity characterizing the contribution from each sublattice, the quantity $\theta_{\mathrm{F}}$ may be called the Faraday rotation angle {\it per sublattice}, or more generally, the one-magnon scattering coefficient.


\end{document}